\documentclass[sigconf, nonacm]{acmart}
\usepackage{hyperref} 
\usepackage{graphicx}
\usepackage{amsmath}
\usepackage{amsthm}
\usepackage{amsfonts}
\usepackage{subfigure}
\usepackage{subcaption}
\usepackage{balance}
\usepackage[ruled,vlined]{algorithm2e}
\usepackage{algpseudocode}
\usepackage{multirow}
\usepackage{xspace}
\usepackage{enumitem}

\newtheorem{theor}{Theorem}

\newtheorem{lem}{Lemma}

\newtheorem{defn}{Definition}

\newcommand{\squishlist}{
 \begin{list}{$\bullet$}
  {  \setlength{\itemsep}{0pt}
     \setlength{\parsep}{3pt}
     \setlength{\topsep}{3pt}
     \setlength{\partopsep}{0pt}
     \setlength{\leftmargin}{2em}
     \setlength{\labelwidth}{1.5em}
     \setlength{\labelsep}{0.5em}
} }
\newcommand{\squishlisttight}{
 \begin{list}{$\bullet$}
  { \setlength{\itemsep}{0pt}
    \setlength{\parsep}{0pt}
    \setlength{\topsep}{0pt}
    \setlength{\partopsep}{0pt}
    \setlength{\leftmargin}{2em}
    \setlength{\labelwidth}{1.5em}
    \setlength{\labelsep}{0.5em}
} }

\newcommand{\squishdesc}{
 \begin{list}{}
  {  \setlength{\itemsep}{0pt}
     \setlength{\parsep}{3pt}
     \setlength{\topsep}{3pt}
     \setlength{\partopsep}{0pt}
     \setlength{\leftmargin}{1em}
     \setlength{\labelwidth}{1.5em}
     \setlength{\labelsep}{0.5em}
} }

\newcommand{\squishend}{
  \end{list}
}

\newcommand{\eat}[1]{}

\newcommand{\kw}[1]{{\ensuremath {\mathsf{#1}}}\xspace}

\newcommand{\stitle}[1]{\noindent{\bf #1}}

\newcommand{\eetitle}[1]{\noindent{\em\underline{#1}}}

\newcounter{ccc}

\DeclareMathOperator*{\argmin}{arg\,min}
\DeclareMathOperator*{\argmax}{arg\,max}

\newcommand\redout{\bgroup\markoverwith
{\textcolor{red}{\rule[.5ex]{2pt}{2pt}}}\ULon}

\newcommand{\mips}{\kw{MIPS}}
\newcommand{\gnns}{\kw{GNNS}}
\newcommand{\psp}{\kw{PSP}}
\newcommand{\nns}{\kw{NNS}}

\newcommand\vldbdoi{10.14778/3725688.3725705}
\newcommand\vldbpages{1770 - 1783}
\newcommand\vldbvolume{18}
\newcommand\vldbissue{6}
\newcommand\vldbyear{2025}
\newcommand\vldbauthors{\authors}
\newcommand\vldbtitle{\shorttitle} 
\newcommand\vldbavailabilityurl{URL_TO_YOUR_ARTIFACTS}
\newcommand\vldbpagestyle{empty} 

\begin{document}
\title{Maximum Inner Product is Query-Scaled Nearest Neighbor}

\author{Tingyang Chen}
\orcid{0009-0008-5635-9326}
\affiliation{%
  \institution{Zhejiang University}
}
\email{chenty@zju.edu.cn}

\author{Cong Fu}
\orcid{0000-0002-3624-6665}
\affiliation{
  \institution{Shopee Pte. Ltd.}
}
\email{fc731097343@gmail.com}

\author{Kun Wang}
\orcid{0009-0009-3950-8809}
\affiliation{
  \institution{Shopee Pte. Ltd.}
}
\email{wk1135256721@gmail.com}

\author{Xiangyu Ke}
\orcid{0000-0001-8082-7398}
\affiliation{
  \institution{Zhejiang University}
}
\email{xiangyu.ke@zju.edu.cn}

\author{Yunjun Gao}
\orcid{0000-0003-3816-8450}
\affiliation{
  \institution{Zhejiang University}
}
\email{gaoyj@zju.edu.cn}

\author{Wenchao Zhou}
\orcid{0009-0002-2689-6020}
\affiliation{
  \institution{Alibaba Group}
}
\email{zwc231487@alibaba-inc.com}

\author{Yabo Ni}
\orcid{0000-0002-7535-8125}
\affiliation{
  \institution{Nanyang Technological University}
}
\email{yabo001@e.ntu.edu.sg}

\author{Anxiang Zeng}
\orcid{0000-0003-3869-5357}
\affiliation{
  \institution{Nanyang Technological University}
}
\email{zeng0118@ntu.edu.sg}

\begin{abstract}
Maximum Inner Product Search (\mips) for high-dimensional vectors is pivotal across databases, information retrieval, and artificial intelligence. 
Existing methods either reduce \mips to Nearest Neighbor Search (\nns) while suffering from harmful vector space transformations, or attempt to tackle \mips directly but struggle to mitigate redundant computations due to the absence of the triangle inequality. 
This paper presents a novel theoretical framework that equates \mips with \nns without requiring space transformation, thereby allowing us to leverage advanced graph-based indices for \nns and efficient edge pruning strategies, significantly reducing unnecessary computations. 
Despite a strong baseline set by our theoretical analysis, we identify and address two persistent challenges to further refine our method: the introduction of the \underline{\textbf{P}}roximity Graph with \underline{\textbf{S}}pherical \underline{\textbf{P}}athway (\psp), designed to mitigate the issue of \mips solutions clustering around large-norm vectors, and the implementation of \underline{\textbf{A}}daptive \underline{\textbf{E}}arly \underline{\textbf{T}}ermination (AET), which efficiently curtails the excessive exploration once an accuracy bottleneck is reached.
Extensive experiments reveal that our method is superior to existing state-of-the-art techniques in search efficiency, scalability, and practical applicability. Compared with state-of-the-art graph-based methods, it achieves an average 35\% speed-up in query processing and a 3$\times$ reduction in index size. 
Notably, our approach has been validated and deployed in the search engines of Shopee, a well-known online shopping platform. Our code and an industrial-scale dataset for offline evaluation will also be released to address the absence of e-commerce data in public benchmarks.

\end{abstract}

\maketitle

\pagestyle{\vldbpagestyle}
\begingroup\small\noindent\raggedright\textbf{PVLDB Reference Format:}\\
\vldbauthors. \vldbtitle. PVLDB, \vldbvolume(\vldbissue): \vldbpages, \vldbyear.\\
\href{https://doi.org/\vldbdoi}{doi:\vldbdoi}
\endgroup
\begingroup
\renewcommand\thefootnote{}\footnote{\noindent
This work is licensed under the Creative Commons BY-NC-ND 4.0 International License. Visit \url{https://creativecommons.org/licenses/by-nc-nd/4.0/} to view a copy of this license. For any use beyond those covered by this license, obtain permission by emailing \href{mailto:info@vldb.org}{info@vldb.org}. Copyright is held by the owner/author(s). Publication rights licensed to the VLDB Endowment. \\
\raggedright Proceedings of the VLDB Endowment, Vol. \vldbvolume, No. \vldbissue\ %
ISSN 2150-8097. \\
\href{https://doi.org/\vldbdoi}{doi:\vldbdoi} \\
}\addtocounter{footnote}{-1}\endgroup

\ifdefempty{\vldbavailabilityurl}{https://github.com/ZJU-DAILY/PSP}{
\vspace{.3cm}
\begingroup\small\noindent\raggedright\textbf{PVLDB Artifact Availability:}\\
The source code, data, and/or other artifacts have been made available at \url{https://github.com/ZJU-DAILY/PSP}.
\endgroup
}

\section{Introduction}
\label{sec:intro}

Maximum Inner Product Search (\mips) is essential across various artificial intelligence and information retrieval applications~\cite{asai2023retrieval, ZhouHLWL22,seo2019real, lewis2020retrieval}. 
The demand for managing large-scale, high-dimensional data — fueled by advancements of large language models~\cite{zhao2023survey} and retrieval-augmented generation~\cite{asai2023retrieval} — has garnered significant attention within the database community~\cite{guo2022manu,pan2023survey,yang2020pase}. 
However, efficient accurate \mips remains a formidable challenge~\cite{teflioudi2015lemp,indyk1998approximate, weber1998quantitative, chen2020hardness}. In response, there has been a shift towards approximate \mips, which trades minimum accuracy for substantial gain in speed.

For the approximate \mips problem, two primary paradigms have emerged. 
The first focuses on the Inner Product (\textsf{IP}) metric, establishing specialized theoretical frameworks to tackle \mips challenges~\cite{morozov2018non,liu2020understanding,tan2021norm,guo2016quantization,zhang2023query,guo2020accelerating,bruch2023approximate}. 
However, the absence of triangle inequality in the \textsf{IP} space hinders them from efficiently reducing redundant computations \cite{tan2021norm}, particularly in recent promising graph-based methods \cite{morozov2018non,liu2020understanding}. 
This deficiency impacts query performance and increases memory usage, as these methods lack theory foundations to prune edges as effectively as advanced \nns graphs \cite{malkov2018efficient,fu2021high,fu2019fast}. 
Specifically, the widely-used greedy graph search algorithm (Algorithm \ref{alg:gnns}) necessitates checking all neighbors at each step to approach the query more closely during traversal, often leading to unnecessary checks that degrade efficiency~\cite{wang2021comprehensive,liu2020understanding,tan2021norm}.
 
The second paradigm addresses the \mips problem by reformulating it as an \nns problem, enabling the use of established \nns methods~\cite{shrivastava2014asymmetric,shrivastava2015improved,neyshabur2015symmetric,li2018general,yan2018norm,zhao2023fargo,zhou2019mobius} that exploit the triangle inequality for efficiency. 
These methods involve various space transformations, which, while effective to some extent, often rely on strong theoretical assumptions~\cite{zhou2019mobius}, can introduce potential structural distortions~\cite{zhao2023fargo, morozov2018non}, and exhibit inefficiencies in data updating (elaborated in \S\ref{sec:preliminaries}).
These significantly restrict their practical applicability.

\begin{figure}[tb!]
\centering
\centerline{\includegraphics[width=0.92\linewidth]{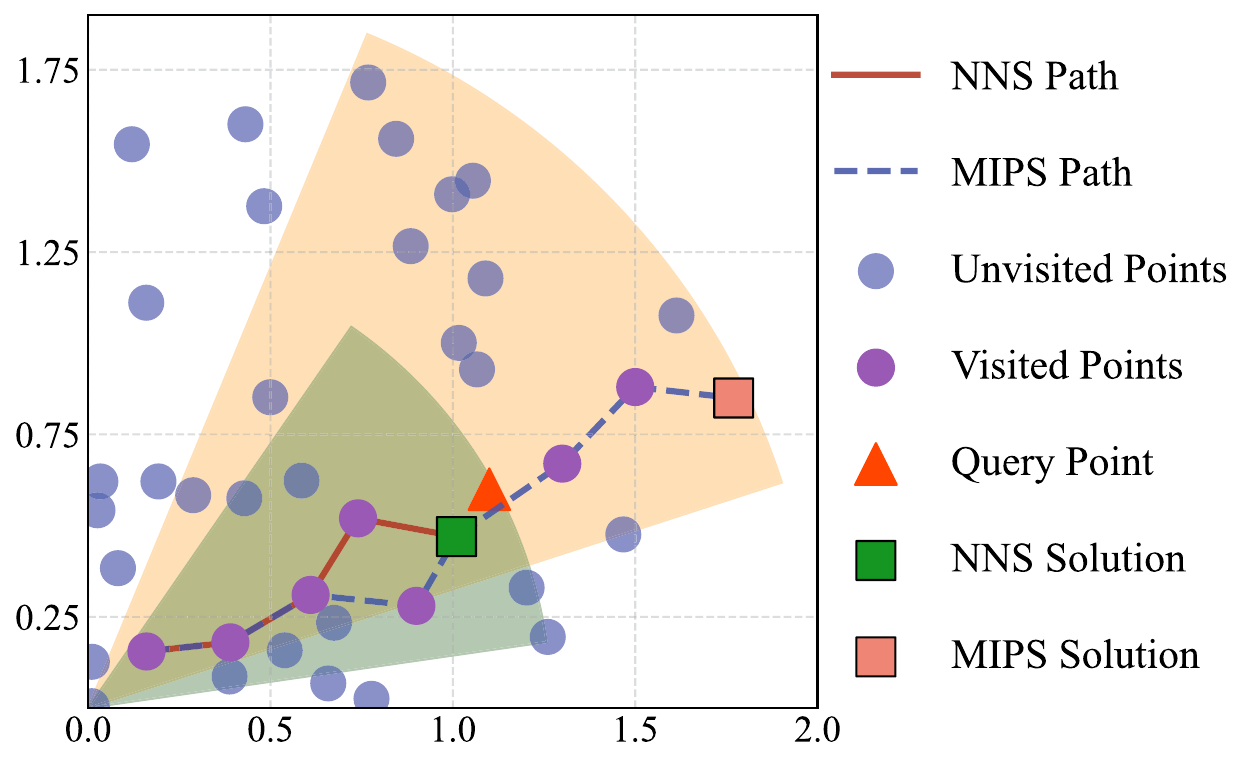}}
\caption{An illustration of the overlaps and discrepancies between the paths of \textsf{NNS} and \textsf{MIPS} on a Euclidean proximity graph. The overlaps hold mainly before the norms of explored nodes exceed the norm of the query.}
\label{fig:diff-search}
\end{figure}

Our research explores a fundamental question: {\em Can we leverage the efficient computation reduction properties without altering the vector space?} 
We begin with an intriguing observation derived from comparing the search routing behaviors of \mips and \nns on the same Euclidean proximity graph, such as an \textsf{MSNET}~\cite{fu2019fast}. 
Specifically, using Algorithm \ref{alg:gnns}, the search iteratively moves from a node to its neighbor, minimizing the distance to the query under given metrics.
As shown in Figure~\ref{fig:diff-search}, the search paths under \mips and \textsf{NNS} {\em initially overlap}. 
The main discrepancy arises when {\em \mips moves beyond the query into outer spherical areas of larger norms}. According to the homogeneity of the {\sf IP} metric, i.e., $\langle \mu q,p \rangle = \mu \langle q,p \rangle$, their paths and solutions coincide when a scalar scales up $q$ until its norm exceeds the \mips solution.

We theoretically verify the above observations by showing that \mips for query $q$ can be equated to \nns for a scaled query $q' = \mu q$ on an \textsf{MSNET}, via introducing a proper scalar $\mu$.
This equality enables using advanced \nns graph indices for \mips without altering the vector space, thus avoiding the associated drawbacks. 
Additionally, this breakthrough lays the groundwork for the first theoretical analysis of search complexity in the graph-based \mips context. 
Our experiments on real-world datasets further validate this theory (\S\ref{sec:exp}). 
Despite these advancements, two practical challenges persist: the tendency of \mips solutions to congregate in large norm regions and the excessive exploration that occurs once a precision bottleneck is reached. 
We propose an innovative index, Proximity Graph with Spherical Pathway (\psp), to handle the biased distribution of \mips solutions. Additionally, we introduce a novel Adaptive Early Termination (\textsf{AET}) mechanism to mitigate excessive explorations.

\noindent{\bf Contributions.} Our contributions are highlighted as follows:

\noindent {\bf (1) Theoretical Foundation:} This work pioneers in establishing comprehensive theoretical foundations for graph-based \mips.

\noindent {\bf (2) Scalable Solution:} We introduce \psp, a scalable graph index with spherical pathways and adaptive search initialization for the biased distribution of \mips solutions. We propose an adaptive early termination mechanism to mitigate the excessive exploration.

\noindent {\bf (3) Extensive Experiments} on 8 real-world datasets validate \textsf{PSP}'s theoretical soundness, efficiency, and scalability, highlighting an average 35\% speed-up in query process and a 3$\times$ reduction in index size over leading graph methods. \textsf{PSP} is the only one reaching 95\% recall under 5ms and thus has been deployed in the large-scale search engines of Shopee, which showcases the application viability.

\noindent {\bf (4) New Industry-scale Dataset:} We release a new dataset derived from industry-scale traffic data collected from the Shopee search scenario. Our dataset fills the absence of e-commercial modality  in public benchmarks, facilitating research in \mips and \nns.

\noindent {\bf Roadmap.} \S\ref{sec:preliminaries} introduces the background of the \mips problem and two main paradigms; \S\ref{sec:analysis} presents our theoretical foundations; \S\ref{sec:method} introduces practical solutions (\psp and \textsf{AET}); \S\ref{sec:exp} outlines the research questions and provides a comprehensive analysis based on the experimental outcomes; \S\ref{sec:discuss-limit} discusses the new dataset and limitations; \S\ref{sec:related} reviews previous related works for broader interests and \S\ref{sec:concl} concludes this paper.

\section{Preliminaries and Background}
\label{sec:preliminaries}
\noindent \textbf{Notations.} $\mathbb{R}^d$ denotes the $d$-dimensional real space. $G=(V,E)$ denotes a directed graph with nodes $V$ and edges $E$. $\delta(\cdot,\cdot)$ denotes the $L_2$ vector distance. $\langle \cdot,\cdot\rangle$ denotes the vector inner product. $\lVert\cdot\rVert$ denotes the $L_2$ vector norm. $sup(\cdot)$ denotes the upper bound.

The Inner Product (\textsf{IP}) serves as a fundamental and effective similarity metric, widely used in artificial intelligence and machine learning \cite{asai2023retrieval,radford2021learning}. With the proliferation of data represented and stored in high-dimensional vectors~\cite{guo2022manu,pan2023survey,zhao2023survey}, \textsf{MIPS} has become increasingly important~\cite{zhao2023fargo,guo2020accelerating,morozov2018non}. \textsf{MIPS} is formally defined as:

\begin{defn}[Maximum Inner Product Search (\textsf{MIPS})]
Given a query $q \in \mathbb{R}^d$, and a dataset $\mathcal{D} \subset \mathbb{R}^d$, the \mips problem aims to find a vector $p \in \mathcal{D}$ that maximizes the inner product with $q$:
 \begin{equation}
     p = \argmax_{p\in \mathcal{D}} \langle p, q \rangle 
 \end{equation}
\end{defn}

Several approaches have been proposed to solve the \textsf{MIPS} problem in sublinear time~\cite{abuzaid2019index,teflioudi2016exact,li2017fexipro}. 
However, they suffer from inefficiencies in query process. Consequently, researchers have explored approximate \mips methods, which trade a minimum accuracy loss for significantly improved retrieval efficiency. Formally, we have:

\begin{defn}[$\epsilon$-Maximum Inner Product Search ($\epsilon$\textsf{MIPS})]
Given a query $q \in \mathbb{R}^d$, a dataset $\mathcal{D} \subset \mathbb{R}^d$, and an approximation ratio $\epsilon \in (0,1)$, let $p^* \in \mathcal{D}$ be the \mips solution of q, the $\epsilon$\mips problem aims to find a vector $p \in \mathcal{D}$ satisfying $\langle p, q \rangle \ge \epsilon \cdot \langle p^*, q \rangle$. 

\end{defn}

While defining $k$-\mips and $k$-$\epsilon$\mips to find the top $k$ solutions is straightforward, we omit these details here for brevity. 
Note that the main difference between \textsf{MIPS} and a similar field, Nearest Neighbor Search (\textsf{NNS}), lies in the metric used \cite{pan2023survey} (inner product and Euclidean distance respectively)
Recent efforts aim to address both $\epsilon${\sf NNS} and $\epsilon$\mips problems with similar strategies: reducing the cost of distance calculations and pruning the search space \cite{zhao2023fargo,ghojogh2018tree,guo2020accelerating,zheng2016lazylsh}. To better motivate our study, we revisit the $\epsilon$\mips techniques from a novel view: {\em whether transforming the vector space}. 

\noindent \textbf{Non-transformation-based methods} use the Inner Product (\textsf{IP}) for both index structures and search algorithms~\cite{morozov2018non,liu2020understanding,tan2021norm,guo2016quantization,zhang2023query,guo2020accelerating}. 
While these methods preserve the integrity of information in the original space, they often suffer from large indices and inferior search performance due to their inability to reduce redundant computations effectively. 
This inefficiency primarily stems from the absence of properties analogous to the triangle inequality in the Euclidean metric. 
In contrast, the triangle inequality significantly reduces the search space in $\epsilon${\sf NNS} under the Euclidean metric \cite{wang2011fast}.

\noindent \textbf{Transformation-based methods}, inspired by the successes in \textsf{NNS}, attempt to reframe the \textsf{MIPS} problem into an \textsf{NNS} problem through two typical space transformation techniques: 

M{\"o}bius transformation~\cite{zhou2019mobius} aims to establish an isomorphism between the Delaunay graph for the \textsf{IP} metric~\cite{morozov2018non} and that for the Euclidean metric~\cite{malkov2014approximate} by scaling the vectors:
\begin{defn}[M{\"o}bius Transformation]
    Given a database $\mathcal{D} \subset \mathbb{R}^d \backslash \{0\}$, the m{\"o}bius transformation modifies $p\in \mathcal{D}$ by $p / \lVert p \rVert^2$.
\end{defn}
\noindent The isomorphism established by M{\"o}bius transformation comes with strong assumptions, such as requiring the origin point to be within $\mathcal{D}$'s convex hull and the data to be independently and identically distributed (\textsf{IID}), which are often violated in practice. The origin-contained assumption could easily be violated with one-hot or multi-hot representations. Verifying that the origin is not contained in such convex hull is straightforward. Additionally, the \textsf{IID} assumption is often violated because the representations generated by \textsf{ML} models are usually correlated in different dimensions~\cite{radford2021learning,schuhmann2022laion}.

XBOX transformation~\cite{zhao2023fargo} adopts an asymmetric approach by elevating both the dataset and query vectors to higher dimensions:
\begin{defn}[XBOX Transformation]
    Given a database $\mathcal{D} \subset \mathbb{R}^d$, and a query $q$. The XBOX transformation maps $\forall p \in \mathcal{D}$ to $p'=\left[p;\sqrt{M^2-\lVert p \rVert^2}\right]$, and maps $\forall q\in\mathcal{R}^d$ to $q'=\left[q;0\right]$:
    where $[;]$ denotes vector concatenation and $M \geq \max_{p\in \mathcal{D}} ||p||$. 
\end{defn}
\noindent Thus, we have $\langle p,q\rangle =\frac{1}{2}\left( \lVert q \rVert^2+ M^2-\delta(p',q') \right)$, a conversion from IP to Euclidean with constant offset. Although this transformation is widely used~\cite{neyshabur2015symmetric,keivani2018improved,pham2021simple,yan2018norm,huang2018accurate}, it has significant drawbacks like potential structural distortions~\cite{morozov2018non} and difficulties in determining an appropriate $M$: A large $M$ may reduce the distinguishability of different points because it contributes significantly to $\delta(P(p),Q(q))$. Conversely, a small $M$ may cause inflexibility in data updates when the new points' norms exceed $M$.

\noindent \textbf{Graph-based methods} for \mips is the focus of this paper. While previous approaches vary in their graph index structures, they predominantly employ a similar search algorithm, often referred to as a greedy walk (Algorithm \ref{alg:gnns}). This algorithm iteratively moves to a neighbor of the current node that is closer to the query, with the search complexity dominated by the \textbf{walking steps and the graph's average out-degree \cite{fu2019fast}}. Notably, non-transformation graph-based methods have struggled with implementing efficient edge pruning strategies due to the absence of the triangle inequality, leading to graphs with high average degrees and subsequently increased search complexity. 
In contrast, advanced graph-based \nns methods \cite{malkov2018efficient,fu2021high,fu2019fast} can dramatically reduce edge quantity while maintaining graph connectivity and short search paths, significantly enhancing efficiency. In the following section, this paper will explore how to leverage these properties for the \mips problem by providing theoretical foundations to align graph-based \mips with \nns~\textbf{without vector space transformations}.

\setlength{\textfloatsep}{0pt}
\begin{algorithm}[t]

  \caption{\textsc{Greedy Search For Graphs\cite{prokhorenkova2020graph}}}
  \label{alg:gnns}
  \LinesNumbered
  \KwIn{Dataset $\mathcal{D}$, Graph $G$, query $q$, candidate set size $l_s$, result set size $k$.}
  \KwOut{Top $k$ result set $R$.}
  $R \gets \emptyset$; $q' \gets \mu q$; $Q \gets \emptyset$;\\
  $P \gets$ random sample $l_s$ nodes from $G$; \\
  \For{each node $p$ in $P$}{
    $Q$.add($p$,$\delta(p, q')$);
  }
  $Q$.make\_min\_heap(); $R$.init\_max\_heap() \Comment{compare distances}\\ 
  \While{$Q$.size()}{
    $p\gets Q.pop()[0]$; $R.insert((p, \delta(p, q')))$\\
    \If{visited($p$)}{continue;}
    $N_p \gets$ neighbors of $p$ in $G$ \\
    $Q\gets$ batch\_insert($Q,N_p$) \\
    $Q.resize(l_s)$;$R.resize(k)$ \Comment{delete larger-distance nodes}\\
  }

  \textbf{return} $R$
\end{algorithm}
\setlength{\textfloatsep}{12pt plus 2pt minus 2pt}

\section{Theoretical Foundations}
\label{sec:analysis}

This section aims to {\em equate \mips and \nns theoretically on Euclidean proximity graphs}. This alignment allows us to capitalize on the beneficial features of established graph-based \nns methods, such as efficient edge pruning, strong connectivity, and robust theoretical guarantees, to enhance graph-based \mips and analyze its search behavior. Formally, a proximity graph is defined as:
\begin{defn}[Proximity Graph]
    Given a dataset $\mathcal{D}$, a proximity graph $G$ on 
    $\mathcal{D}$ can be denoted by $G=(V,E)$, where $V$ is the node set, and $E$ is the edge set. Each node $v_i \in V$ represents a vector $p_i\in \mathcal{D}$, while $(v_i,v_j) \in E$ if and only if $(v_i,v_j)$ satisfies a given criteria. 
\end{defn}

Within our theoretical alignment, the specific type of proximity graph—whether a Delaunay graph \cite{aurenhammer1991voronoi}, a Navigating Small World (\textsf{NSW}) graph \cite{malkov2018efficient}, or a Monotonic Relative Neighborhood Graph (\textsf{MRNG}) \cite{fu2021high}—is not fixed. {\em The choice of graph does not affect the validity of our theory.} 
Notably, part of our theory also is {\em not limited to graph-based context} (Theorem \ref{theorem:global}) and can be applied to any type of \mips methods, such as hashing and quantization.

\subsection{Equivalence under Scaled Query}
\label{sec:scale}

First, we establish a mapping $q'=f(q)$, such that {\bf (1)} the \mips solutions for $q'$ align with those for $q$, and {\bf (2)} the \nns solutions for $q'$ align with the \mips solutions for $q$. This mapping enables the use of efficient {\sf NNS} algorithms to solve the \mips problem without changing the origin vector space. We leverage the homogeneity of {\sf IP} metric, as formalized in the following lemma:

\begin{lem}
\label{lem:lem-1}
    Given a vector database $\mathcal{D} \subset \mathbb{R}^d$, and a scalar $\mu > 0$, the \mips solution for $q$ is identical to those for $q' = \mu q$ within $\mathcal{D}$.
\end{lem}

\begin{proof}
    Let $p^* \in \mathcal{D}$ be a \textsf{MIPS} solutions for $q$, satisfying $\forall p \in \mathcal{D}\backslash\{p^*\}, \langle p^* , q \rangle \ge \langle p, q \rangle$. Scaling $q$ by $\mu >0$, we have:
    \begin{equation}
        \langle p^* ,\mu q \rangle = \mu \langle p^*,q \rangle \ge  \mu\langle p,q\rangle=\langle p, \mu q \rangle, \forall p \in \mathcal{D}\backslash\{p^*\}
    \end{equation}
    Therefore, $q'=\mu q$ and $q$ share the same \mips solutions. 
\end{proof}

Lemma \ref{lem:lem-1} reveals that non-negative constant scaling of the query vector preserves \mips solutions. Next, we will align \mips and \nns with this scaling mapping by identifying solution space for $\mu$.

\begin{theor}
    \label{theorem:global}
    Given a vector database $\mathcal{D} \subset \mathbb{R}^d$ and $\forall q \in \mathbb{R}^d$, there exists a scalar $\bar{\mu}$ such that for $\forall \mu > \max(\bar{\mu},0)$, the nearest neighbor of $q'=\mu q$ in $\mathcal{D}$ aligns with the \mips solution for $q$.
\end{theor}

\begin{proof}
We exclude cases where the zero-vector or multiple identical vectors in $\mathcal{D}$ complicate the analysis. Then, we break down the proof step by step.


\textbf{Case 1: } Assume there is a unique \mips solution $p^* \in \mathcal{D}$ for $q$. By Lemma \ref{lem:lem-1}, $p^*$ is also a \mips solution for $q', \forall \mu>0$.

To ensure $p^*$ is a nearest neighbor of $q' = \mu q$, we need to solve:
\begin{equation}
    \lVert p^* -q' \rVert  < \lVert p-q' \rVert, \quad \forall p \in \mathcal{D}, p \neq p^* \\ \nonumber
\end{equation}
By simplifying and rearranging, we find the condition for $\mu$:
\begin{equation}
    \mu >  \frac{\lVert p^*\rVert{^2} - \lVert p\rVert^2}{2(\langle p^*, q \rangle -  \langle p, q \rangle)}, \forall p\in \mathcal{D}, p\neq p^*
\end{equation}

Let $\bar{\mu} =\sup \left(\left\{\frac{\lVert p^*\rVert{^2} - \lVert p\rVert^2}{2(\langle p^*, q \rangle -  \langle p, q \rangle)}\Big| \forall p\in \mathcal{D}, p\neq p^*\right\}\right)$. This space is bounded because $\mathcal{D}$ is a finite set; thus, $\bar{\mu}$ exists. Unifying conditions for $\mu$, $p^*$ is a nearest neighbor of $q'$ when $\mu > \max(\bar{\mu},0)$.

\textbf{Case 2: } Considering there may exist multiple \mips solutions for $q \in \mathcal{D}$, let $\mathcal{P}^* = \left\{p^*|\langle p^*, q \rangle \ge \langle p, q \rangle, \forall p \in \mathcal{D}, p \neq p^*\right\}$ denote this solution set. We can find a $p_n^* \in \mathcal{P}^*$ such that $p_n^*$ is the nearest neighbor of $q'=\mu q$ with an appropriate $\mu$. By following a similar procedure in Case 1, we can get the solution space for $\mu$ as:
\begin{equation}
 \mu > \max(\bar{\mu},0) 
 = \max (\sup \left(\left\{\frac{\lVert p^*\rVert{^2} - \lVert p\rVert^2}{2(\langle p^*, q \rangle -  2\langle p, q \rangle)}|p \in \mathcal{D} \setminus \mathcal{P}^* \right\} \right),0) \nonumber
\end{equation}

Summarizing Case 1 and 2, we can identify a scalar boundary $\bar{\mu}$ such that $\forall \mu > \max(\bar{\mu},0)$, there exists a point $p^*\in \mathcal{D}$ which is the nearest neighbor of $q'=\mu q$ and also a \mips solution of $q$.
\end{proof}

By Theorem \ref{theorem:global}, we establish the existence of a scalar $\mu$ and a corresponding node $p^*$ that {\em unifies the solutions for the \nns and \mips problems} for a scaled query $q'=\mu q$. This identification alone does not ensure we can retrieve $p^*$ under the \textsf{IP} metric as efficiently as \nns algorithms on a Euclidean proximity graph. To address this, we aim to demonstrate {\em the equivalence of search behavior} under \textsf{IP} and Euclidean metrics within the same graph structure:

\begin{theor}
\label{theorem:local}
Given a proximity graph $G=(V,E)$ and a query $q \in \mathbb{R}^d$, when using the standard Graph Nearest Neighbor Search (\gnns)~\cite{prokhorenkova2020graph} algorithm to decide the \mips solution for $q$, there exists a scalar $\bar{\mu}$ such that for $\forall \mu > \max(\bar{\mu},0)$, we can identify a node $p^* \in \mathcal{N}_o$ that satisfies: $p^* \in \left\{\argmax_{p \in \mathcal{N}_o}  \langle p, q \rangle\right\} \cap \left\{\argmin_{p \in \mathcal{N}_o} \delta(p, q')\right\}$. Here, $o$ can be any node on the search path of \gnns regarding the query $q'=\mu q$, and $\mathcal{N}_o=\left\{p|(p,o) \in E \right\}$ denotes $o$'s neighbor nodes.  
\end{theor}

\begin{proof}[Proof Sketch]
To align the search paths under both metrics as per Algorithm~\ref{alg:gnns}, we can unify the node-selection behavior by localizing the problem: solving for the optimal $p^*$ among the neighbors at each greedy step, akin to the conditions established in Theorem~\ref{theorem:global}. Let $l$ be the number of steps in the path. Define $U=\left\{\bar{\mu}_i| 1 \le i \le l\right\}$. it's not hard to verify that there  exists a scalar boundary $\sup \left(U\cup\{0\} \right)$ forces the Algorithm \ref{alg:gnns} under both \textsf{IP} and Euclidean metrics to generate the same search path targeting query $\mu q$. Please refer to \cite{chen2025maximum} for the complete proof.
\end{proof}

\noindent{\bf Remarks.} From Theorem \ref{theorem:global} and Theorem \ref{theorem:local}, we discern a compelling duality between the \mips and \nns paradigms. These findings suggest that by appropriately scaling the query $q$, we can modulate the overlap in their search paths on a Euclidean proximity graph. 
{\em By choosing an appropriate $\mu$, the search behavior becomes invariant to the metric used.} 
Our experimental validations further corroborate this theoretical insight on real-world datasets (\S\ref{sec:exp-result}).

Notably, in practice, it is unnecessary to find a specific $\mu$ to tackle the \mips problem (Lemma \ref{lem:lem-1}). 
Theorems \ref{theorem:global} and \ref{theorem:local} demonstrate that the {\sf IP} metric can be directly applied on a Euclidean proximity graph to perform IP-greedy search with Algorithm \ref{alg:gnns}. 

\subsection{Efficiency Analyses for Graph-Based MIPS}
\label{sec:num-stable}
While our theoretical framework permits the use of any Euclidean Proximity Graph, only a few, such as \textsf{MRNG} \cite{fu2019fast} and \textsf{SSG} \cite{fu2021high}, come with robust theoretical guarantees. 
To overcome the high-degree issue prevalent in non-transformation graph-based methods \cite{morozov2018non,liu2020understanding}, we aim to identify a sparse Euclidean graph that ensures a low amortized search complexity. 
In this paper, we focus on \textsf{SSG} for its sparsity and additional theoretical guarantees for queries absent from the database $\mathcal{D}$. 

\noindent \textbf{1-MIPS Analysis.} We begin with top-1 \mips of \textsf{SSG} formally as:

\begin{theor}
\label{theorem:path-len}
    Consider a vector database $\mathcal{D} \subset \mathbb{R}^d$ containing $n$ points, where $n$ is sufficiently large to ensure robust statistical properties. 
    Let $\mathcal{G}$ be a SSG \cite{fu2021high} defined on $\mathcal{D}$. 
    Assume that the base and query vectors are independently and identically distributed (i.i.d.) and are drawn from the same Gaussian distribution, with each component having zero mean and a variance $\sigma^2$. 
    The expected length of the search path $L$ from any randomly selected start node $p$ to a query $q\in \mathbb{R}^d$ can be bounded by 
    $\mathbb{E}[L]<c_0\frac{\log n + c_1d}{\log R + c_2d}$,
    where $R$ is the max-degree of all possible $\mathcal{G}$, independent with $n$ \cite{fu2021high}, and $c_0,c_1,c_2$ are constants.
\end{theor}

\begin{proof}[Proof Sketch]
Given Theorem \ref{theorem:global}, \ref{theorem:local}, along with the monotonic search property of \textsf{SSG} \cite{fu2021high}, Algorithm \ref{alg:gnns} identifies a greedy search path where each step minimizes the Euclidean distance to $\mu q$ while maximizes the {\sf IP} distance with respect to $q$, i.e. $\langle r_i, q\rangle > \langle r_{i-1}, q\rangle$. The expected length of the search path can be calculated as:
\begin{equation}
    \mathbb{E}[L] = \mathbb{E}_{p\in \mathcal{D}, q \in \mathbb{R}^d}\left[\frac{\langle r_{mip}, q\rangle - \langle p, q \rangle}{\mathbb{E}[\langle r_i, q \rangle - \langle r_{i-1}, q \rangle |r_{i-1}, q]}\right]
\end{equation}
where $r_{mip}$ is the \mips solution of $q$. 
The outer expectation cannot be simplified directly due to potential dependencies among $p$, $r_i$, and $q$. 
Given that the base and query vectors are finite and drawn from the same distribution, we can enclose them in a hypersphere of diameter $2H$~\cite{weisstein2002hypersphere}, which is independent of $p$, $r_i$, and $q$. 
Since $\langle p, q \rangle < H^2$ and $\langle r_{mip}, q \rangle < H^2$, we can derive:
\begin{equation}
    \label{eq:ubl}
    \mathbb{E}[L] < \frac{2H^2}{\mathbb{E}_{p\in \mathcal{D}, q \in \mathbb{R}^d}\left[\mathbb{E}\left[\langle r_i, q \rangle - \langle r_{i-1}, q \rangle |r_{i-1}, q\right]\right]}
\end{equation}

Although the exact solution to this inequality is intractable, a practical approach involves approximating both the numerator and the denominator. 
By applying Extreme Value Theory ({\sf EVT})~\cite{smith1990extreme}, we can derive a tight upper bound for this approximation, ensuring convergence to the bound as $n$ becomes sufficiently large.

As $H$ is the half-diameter, $H^2$ is the maximum squared norm of the dataset $\mathcal{D}$.
We define $M_0 = H^2 = \max_1^n\{X_1^2, ..., X_n^2\}$, where $X_i$ are element-wise i.i.d. samples from $\mathcal{N}(0,\sigma^2)$.
Consequently, $X^2$ follows a chi-square distribution with $d$ degrees of freedom. 
The Moment Generating Function ({\sf MGF})~\cite{curtiss1942note} of $M_0$ is given:
\begin{equation}
    MGF(t)_{X^2} = (1-2\sigma^2t)^{-\frac{d}{2}}, 0< t < \frac{1}{2\sigma^2}
\end{equation}
From Jensen's Inequality~\cite{mcshane1937jensen} with $\phi(x)=e^{t_0x}, t_0>0$, we have:
\begin{equation}
    \begin{aligned}
        e^{t_0\mathbb{E}[M_0]} &\le \mathbb{E}[e^{t_0M_0}] = \mathbb{E}\max_{i=1}^ne^{t_0X_i^2} \\
        &\le \sum_1^n\mathbb{E}\left[e^{t_0X_i^2}\right] = nMGF(t_0)_{X^2} 
    \end{aligned}
\end{equation}
\begin{equation}
    \mathbb{E}[M_0] \le \frac{1}{t_0}\left(\log n + \frac{d}{2}\log (1-2\sigma^2t_0)^{-1}\right) 
\end{equation}
Here, $t_0$ is a hyper-parameter that can be optimized within the range $(0, 0.5)$ to enhance the tightness of this bound. 

A similar approach can be applied to the random variable $XY$ for approximating the denominator of Eq.~\ref{eq:ubl}, where $X$ and $Y$ element-wise i.i.d. sampled from $\mathcal{N}(0,\sigma^2)$. 
In this case, $XY$ follows a distribution characterized by a modified Bessel function of the second kind~\cite{bowman1958introduction}. Then we can have:
\begin{equation}
    \mathbb{E}[\langle r_i,q \rangle] \le \mathbb{E}[\langle r_{i-1}, q \rangle] + \frac{1}{t_1}\left(\log R + d \log (1-\sigma^2t_1^2)^{-1}\right)
\end{equation}

Substituting them into Eq.(\ref{eq:ubl}), we have:
\begin{equation}
\begin{aligned}
    \mathbb{E}[L] &< \frac{2H^2}{\mathbb{E}_{p\in \mathcal{D}, q \in \mathbb{R}^d}\left[\mathbb{E}\left[\langle r_i, q \rangle - \langle r_{i-1}, q \rangle |r_{i-1}, q\right]\right]} \\
    &\approx \frac{\frac{2}{t_0}\left(\log n + \frac{d}{2}\log (1-2\sigma^2t_0)^{-1}\right)}{\frac{1}{t_1}\left(\log R + d\log (1-\sigma^2t_1^2)^{-1}\right)}
\end{aligned}  
\end{equation}
where $R$ is the maximum degree of $\mathcal{G}$, independent of $n$ and choices of $p$, $q$, and $r_i$\cite{fu2021high}.
Thus, the outer expectation can be eliminated. 
When $t_0$ and $t_1$ are optimized for the tightest approximation, introducing constants $c_0$, $c_1$ simplifies the formula to:
\begin{equation}
\mathbb{E}[L] < c_0\frac{\log n + c_1d}{\log R + c_2d}
\end{equation}

The complete process of derivation can be found in \cite{chen2025maximum}.
\end{proof}

While Theorem \ref{theorem:path-len} provides an approximation for the expected search path length, the results align well with intuitive insights: the length $L$ increases gradually with $n$ while decreases with graph max-degree $R$. 
In \textsf{SSG}, the inter-edge angle $\alpha$ governs the graph sparsity \cite{fu2021high}. 
A smaller $\alpha$ leads to higher $R$, enhancing graph connectivity and thereby reducing inter-node transition costs. 
Moreover, the overall search complexity for top-1 \mips can be bounded by $O\left(c_0R\frac{\log n + c_1d}{\log R + c_2d}\right)$, indicating that the complexity increases more rapidly with $R$ than with $n$, highlighting the critical role of graph pruning for better searching efficiency.

\noindent \textbf{$k$-MIPS Analysis.} The preceding theory examines the validity and efficiency of tackling the 1-\mips problem on an \textsf{SSG}. Next, We demonstrate that high-confidence $k$-\mips solutions can be obtained by exploring the neighborhoods of the 1-\mips solution. 

\begin{theor}
\label{theorem:k-mip-neighbor}
    Given a vector database $\mathcal{D} \subset \mathbb{R}^d$ consisting of $n$ points, where $n$  is sufficiently large to ensure robust statistical properties. For the ease of calculation, we assume base and query vectors are element-wise i.i.d. and are drawn from the same Gaussian distribution, parameterized with zero mean and a variance $\sigma^2$. Given a tunable parameter $s>0$, we have $\langle r, q\rangle$ is lower bounded by $\left|\lVert p\rVert - s\right|\lVert q\rVert\cos{ \left(\theta_{pq} + 2\sin^{-1}\frac{s}{2\lVert p\rVert} \right)}$ with probability $Q(s) = \frac{\gamma \left(\frac{d}{2},\frac{s}{2\sigma^2} \right)}{\Gamma \left(\frac{d}{2} \right)}$, where $\Gamma(\cdot)$ is the gamma function and $\gamma(\cdot)$ is the lower incomplete gamma function.
\end{theor}
\begin{proof}
For any query $q$ and any point $r \in \mathcal{D}$, this proof aims to establish their relationship regarding $q$'s top-1 \mips solution $p$,  We begin by expand $\langle r, q\rangle$ for $r$ and $q$ as:
\begin{equation}
    \langle r, q\rangle = \lVert r\rVert\lVert q\rVert\cos{\theta_{rq}}
\end{equation}
, where $\theta_{rq}$ is the angle between vector $r$ and $q$. Under the same norm of $r$, $\cos{\theta_{rq}}$ w.r.t. $p$ is minimized when normalized $r$, $p$, and $q$ fall on the same great circle of the unit sphere. According to spherical trigonometry cosine rule, above equation is bounded as:
\begin{equation}
    \langle r, q\rangle \ge \lVert r\rVert \lVert q\rVert \cos{(\theta_{rp} + \theta_{pq})}
\end{equation}
With triangle inequality, we further have:
\begin{equation}
    \langle r, q\rangle \ge \left|\lVert p\rVert-\Delta_{pr}\right|\lVert q\rVert \cos{(\theta_{rp} + \theta_{pq})}
\label{eq:ip-lb}
\end{equation}
, where $\Delta_{pr}$ is the $L_2$ distance between point $p$ and $r$. We can bound $\Delta_{pr}$ according to the distribution of $\Delta_{pr}$. Specifically, let $Z=\lVert X-Y\rVert^2$ denote the distribution of Euclidean distance between $X$ and $Y$, i.i.d. drawn from $\mathcal{D}$. $Z$ follows a gamma distribution parameterized as :
\begin{equation}
    Z = \lVert X-Y\rVert^2 \sim Gamma \left(\frac{d}{2},2\sigma^2 \right)
\end{equation}
The probability that $\mathbb{P}[Z < s]$ for a given threshold $s$ can be derived from the Cumulative Density Function (CDF) as:
\begin{equation}
    \mathbb{P}[Z < s] = Q(s) = \frac{\gamma \left(\frac{d}{2},\frac{s}{2\sigma^2} \right)}{\Gamma \left(\frac{d}{2} \right)}
\label{eq:l2-norm-lb}
\end{equation}
With a given $\Delta_{pr}$, the maximum of $\theta_{rp}$ can also be calculated as $\max \theta_{rp} = 2\sin^{-1}\frac{s}{2\lVert p\rVert}$. Combining Equation (\ref{eq:ip-lb}) and (\ref{eq:l2-norm-lb}), we derive the lower bound for $\langle r, q\rangle$ as :
\begin{equation}
    \langle r, q\rangle > \left|\lVert p\rVert - s\right|\lVert q\rVert\cos{\left(\theta_{pq} + 2\sin^{-1}\frac{s}{2\lVert p\rVert}\right)}
\end{equation}
, with probability $Q(s)$ in Equation (\ref{eq:l2-norm-lb}).
\end{proof}

According to Theorem \ref{theorem:k-mip-neighbor}, reducing $s$ drives the convergence of $\langle r, q \rangle$ towards $\langle p, q \rangle$, while also decreasing the estimated number of nodes ($nQ(s)$) that maximize the IP values relative to $q$. 
Since $s$ represents the upper bound of the Euclidean distance between $r$ and $q$, a judicious choice of $s$ can effectively identify about $nQ(s)$ points in $\mathcal{D}$ as both the top-$k$ \mips solutions for $q$ and the range-$s$ nearest neighbors of $p$. 
In most Euclidean proximity graphs, it is likely that neighbors of neighbors are also neighbors \cite{prokhorenkova2020graph}. 
Thus, after locating the top-1 \mips solution $r_0$ for query $q$, Algorithm \ref{alg:gnns} may require extra iterations to navigate among $r_0$'s nearest neighbors for the remaining $k-1$ solutions. 
Due to the efficient pruning in sparse graphs like \textsf{SSG}, the nearest neighbors of $r_0$ are expected to be no more than $O(f(n)n^{1/d}\log n)$ hops away \cite{fu2019fast}, where $f(n)$ is a slowly increasing function of $n$. Thus, the search complexity for top-k \mips is bounded by $O\left(c_0\frac{\log n + c_1d}{\log R + c_2d} + c_3f(n)kRn^{1/d}\log n\right)$, where $c_3$ is a scaling constant that integrates the terms linearly.

\begin{figure}[tb!]
\centering
\centerline{\includegraphics[width=0.95\linewidth]{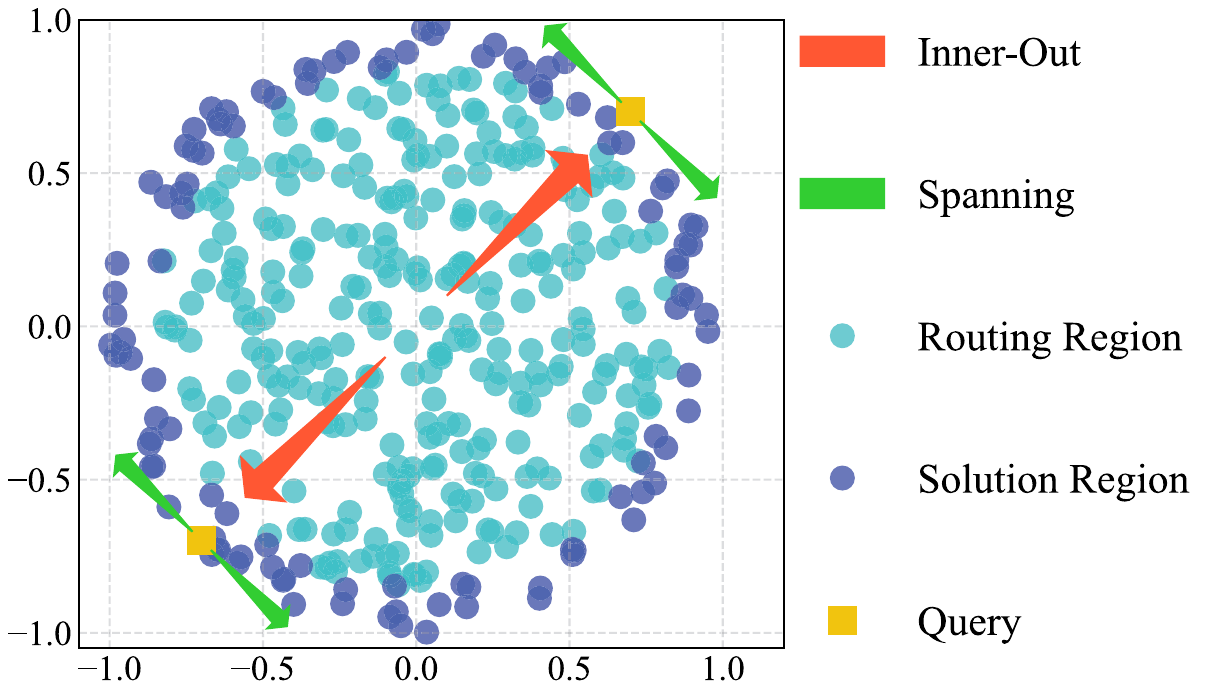}}
\caption{An illustration of biased solutions for \textsf{MIPS} problems. Spherical Pathways benefit the routing from the inner to the outer ring and the spherical spanning for k answers.}
\label{fig:bias-solution}
\end{figure}

\setlength{\textfloatsep}{0pt}
\begin{algorithm}[t!]

  \caption{\textsc{Graph Index Construction}}
    \label{alg:index}
  \LinesNumbered
  \KwIn{Dataset $\mathcal{D}$, candidate size $L$, maximum degree $R$, minimum angle $\alpha$, refine edge quota $S$, navigation sample number $n$, $k$NNG neighbor number $K$.}
  \KwOut{\psp index $G$}

    $G \gets NSSG\_Build(\mathcal{D}, R, \alpha, L,K);$ \Comment{refer to \textsf{SSG} paper~\cite{fu2021high}} \\
    \For{each node $i$ in $G$}{
        $G \gets$ \textsf{EF}$(i,S,\alpha,G)$; \Comment{\textsf{refer to Alg. \hyperref[alg:refine]{2}}} \\
    }
    N $\gets$ \textsf{SN}$(\mathcal{D}, n)$ \Comment{\textsf{refer to Alg. \hyperref[alg:nsn]{3}}} \\
    $G\gets \{G,N\}$ \\
      
  \textbf{return} $G$

\end{algorithm}
\noindent\textbf{Remarks}. Previous transformation-based methods \cite{shrivastava2014asymmetric,shrivastava2015improved,neyshabur2015symmetric,li2018general,yan2018norm,zhao2023fargo,zhou2019mobius} sought to reduce the \mips problem to the \nns problem using non-linear transformations, suffering from strong theoretical assumptions~\cite{zhou2019mobius}, structural distortions~\cite{zhao2023fargo, morozov2018non}, and complicate data updates (\S \ref{sec:preliminaries}), while failed to provide theoretical guarantees on the reachability of the 1-\mips solution. 
Our Theorems \ref{theorem:global} and \ref{theorem:local} demonstrate that this reachability can be achieved without space transformations. 
Theorems \ref{theorem:path-len} and \ref{theorem:k-mip-neighbor} validate the efficiency of solving \mips on an \textsf{SSG}, highlighting the crucial role of edge pruning in search performance — a factor which is previously overlooked\cite{morozov2018non,liu2020understanding}. 
Our extensive experiments in \S \ref{sec:exp-result} substantiate this analysis.

\section{Practical Solution}
\label{sec:method}
In this section, we introduce a novel \mips
framework tailored for practical applications. This framework comprises two key components: a new graph-based indexing structure named \textbf{Proximity Graph with Spherical Pathways (\textsf{PSP})} and a novel search algorithm with \textbf{Adaptive Early Termination (\textsf{AET})}. The \textsf{PSP} index is specifically designed to address the \textbf{Biased Solution} problem prevalent in \mips approaches. Concurrently, the \textsf{AET} mechanism effectively mitigates the \textbf{Excessive Exploration} problem, enhancing search efficiency across diverse query distributions.

\subsection{Proximity Graph with Spherical Pathway}
\label{sec:index}

\noindent {\bf Base Structure.} Advanced theoretical models for nearest neighbor search often exhibit $O(N^2\log N)$ indexing time complexity~\cite{fu2019fast,fu2021high}, which impedes practical application of our theoretical insights. To mitigate this, we employ an approximate Euclidean proximity graph, \textsf{NSSG}, a practical adaptation of its theoretical counterpart \textsf{SSG}~\cite{fu2021high}. This choice is driven by its efficiency and our theoretical analysis in \S\ref{sec:analysis}. Please refer to \textsf{SSG} paper for details in \textsf{NSSG} indexing algorithm.

\setlength{\textfloatsep}{0pt}
\begin{algorithm}[t!]
  \caption{\textsc{Edge Refinement (\textsf{EF})}}
  \label{alg:refine}
  \LinesNumbered
  \KwIn{Node $n$, new edge quota $S$, angle $\alpha$, \textsf{NSSG} $G$.}
  \KwOut{A refined graph $G'$} 
  $E_n \gets \emptyset$; $C \gets$ 2-Hop neighbors of n on $G$; $G'\gets G$\\
  
  sort $C$ in descending order of $\langle n, c_i \rangle$, $c_i \in C$; \\

  $E_n$.add($(n,C[0])$); $C$.remove($C[0]$); $o \gets$ zero vector;\\
  \While{$C$ is not empty} {
    $p \gets$ $C[0]$; $C$.remove($C[0]$);\\
    
   \If{$cos\angle nop > \alpha$}{ 
    continue;
   }
   $E_n$.add($(n, p)$); \\
   \If{$E_n$.size() $\ge S$}{
        break; 
   }  
  }
  \For{each edge $e$ in $E_n$}{
    $G'$.add($e$) \Comment{\textsf{deduplicate if repeated}}
  }
  \textbf{return} $G'$
\end{algorithm}
\setlength{\textfloatsep}{12pt plus 2pt minus 2pt}

\setlength{\textfloatsep}{0pt}
\begin{algorithm}[t!]
  \caption{\textsc{Spherical Navigation (\textsf{SN})}}
  \label{alg:nsn}
  \LinesNumbered
  \KwIn{Dataset $\mathcal{D}$, navigation sample number $m$.}
  \KwOut{Inverted File $N$} 
  $N \gets \emptyset$;$\mathcal{D}'\gets$ Sample($\mathcal{D}$) \Comment{cluster on subset $\mathcal{D}'$ for big $\mathcal{D}$}\\
  $C \gets$ Normalized k-means$(\mathcal{D},\mathcal{D}', c)$;\Comment{\textsf{c is cluster number}} \\
  $m_c \gets m /c $; \\
  \For{$i \in range(0,c)$}{
    $S_m \gets$ sampling $m_c$ nodes from $C[i]$ with Gaussian; \\
    $N[i]$.add$(S_m)$; \\
  }

  \textbf{return} $N$
\end{algorithm}
\setlength{\textfloatsep}{12pt plus 2pt minus 2pt}

\noindent {\bf Proximity Graph with Spherical Pathways (PSP)} addresses the biased solution problem by injecting new edges into the SSG, named as Spherical Pathways (SP), and introducing a lightweight extra structure for search entry management, named as Spherical Navigation (SN). These two components are introduced as follows.

Solving the \mips problem often results in biased solutions towards points in regions of large norms, as highlighted in ~\cite{liu2020understanding}. 
Traditional proximity graphs under the Euclidean metric typically do not inherently address this bias, as their edge selection criteria focus on local neighborhood connectivity, leading to limited sensory regions~\cite{wang2021comprehensive}. 
Empirical observations (Figure \ref{fig:bias-solution}) reveal that \mips routing typically involves {\em initially hopping towards a few answers from inner to outer rings, followed by a broader exploration process for the remaining answers}. To enhance this process, we introduce \textbf{Spherical Pathways}, which benefit the \mips routing in two aspects (Figure \ref{fig:bias-solution}): {\bf (1)} Linking each node to its \mips neighbors in the outer ring to accelerate the inner-out routing; {\bf (2)} Strengthening mutual connectivity within the "outer ring", thereby expanding sensory regions crucial for the spanning process in $k$-$\epsilon$\mips.

Identifying \mips neighbors for each node on a large scale is challenging; we address this with a heuristic that involves collecting $k$-hop neighbors and selecting up to $S$ nodes with the largest \textsf{IP} distances as new neighbors. To widen the sensory region, we introduce a smallest-angle constraint inspired by the techniques used in \textsf{NSSG} indexing~\cite{fu2021high}. Specifically, any new \mips neighbor $m$ of node $n$ (excluding the one with the largest \textsf{IP} distance), the angle $\angle nom \ge \alpha$. Empirically, we find that collecting 2-hop neighbors has significantly enhanced search performance by 8\% (\S \ref{sec:exp-result}). This process, termed \textbf{Edge Refinement (\textsf{EF})}, is detailed in Algorithm \ref{alg:refine}.

\begin{figure}[tb!]
\centering
\centerline{\includegraphics[width=0.92\linewidth]{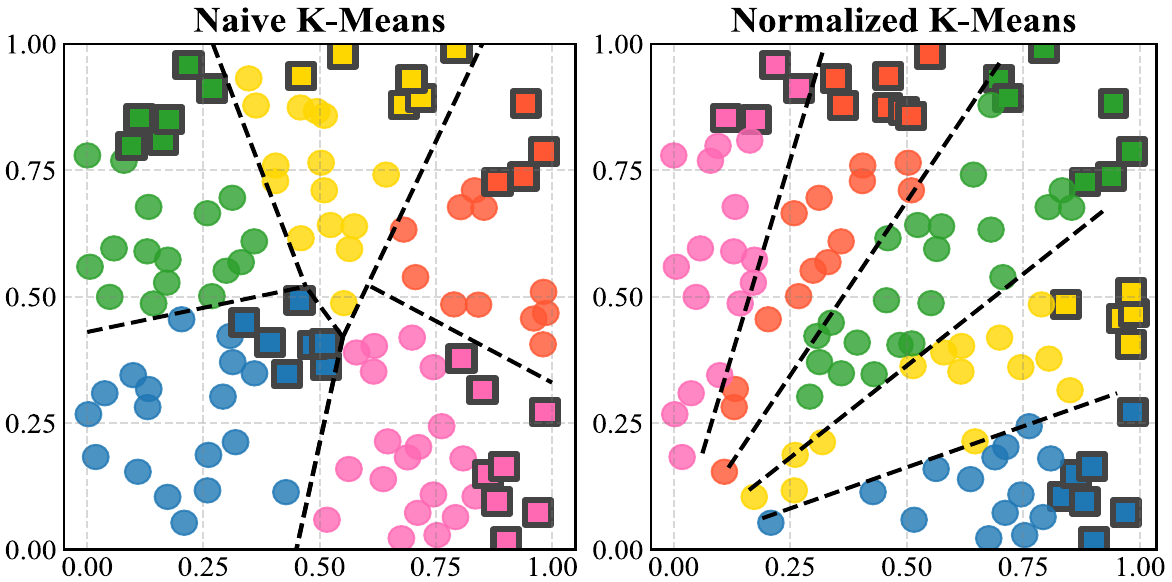}}
\caption{An illustration of k-means-based navigation node selection w/ and w/o normalization. Squared with dark edges are randomly selected navigation nodes with large norms.}
\label{fig:nsn}
\end{figure}

\noindent {\bf Spherical Navigation (\textsf{SN}).} Navigation nodes (entry nodes of the search algorithm) are pivotal in optimizing search performance on proximity graphs~\cite{fu2016efanna,malkov2018efficient}. Given the bias in \mips solutions towards large-norm vectors, selecting navigation nodes among them can be advantageous. To prevent over-concentration of navigation nodes, we cluster the base vectors' normalised projections on a unit sphere using $k$-means, prioritizing angular separation over Euclidean distance (see Figure \ref{fig:nsn}). 
We then sample $m/c$ navigation nodes from each of the $c$ clusters based on their norm distribution in original space. While the norm distribution follows a long-tail pattern, the Gaussian distribution can serve as an effective approximation in practice. 
The resulting structure maintaining the selected navigation nodes is encapsulated as an Inverted File (\textsf{IVF})~\cite{jegou2010product}. Clusters are keyed by their centers, while navigation nodes are stored in corresponding inverted lists. The \textsf{SN} selection process is outlined in Algorithm~\ref{alg:nsn}. Notably, while \textsf{SN} is built on normalised vectors, it only generates the entry points' IDs. IP metric is used in the original space when searching on the graph, aligning with our theory.

To summarize, building a \textsf{PSP} (detailed in Algorithm \ref{alg:index}) involves three steps: (1) Build an \textsf{NSSG}; (2) Apply \textsf{EF}; (3) Apply \textsf{SN}.

\subsection{Searching on PSP}
\label{sec:search}

\begin{figure*}[t!]
	\centering
	\includegraphics[width=0.97\linewidth]{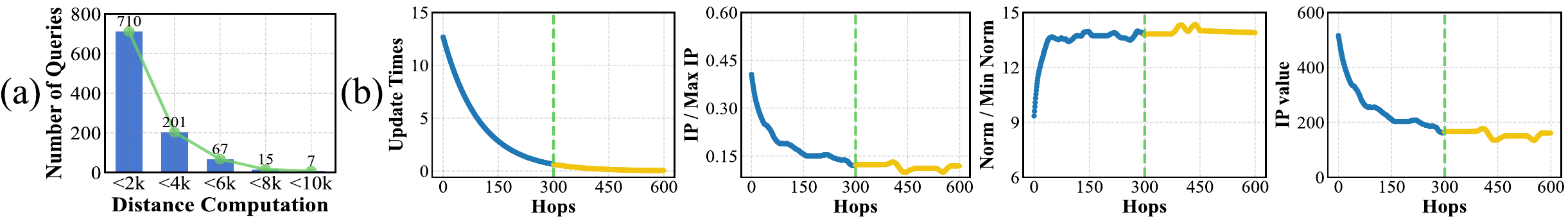}
	\caption{An illustration of early termination intuitions. Figure (a) shows the quantity distribution of visited nodes to reach 99\% recall@100 on \textsf{Text2Image1M}. Figure (b) presents the changing trends of four selected features during search for a query. The green dotted lines indicate the optimal stop state. After that (yellow curves), the recall@100 stops increasing.}
    \label{fig:intuition-early-stop}
\end{figure*}

Standard \textsf{GNNS} \cite{prokhorenkova2020graph} does not fully release the potential of \textsf{PSP}. To optimize its efficiency for \mips and ensure effective alignment with \psp, we employ three key adaptations: {\bf (1)} selecting navigation nodes based on \textsf{SN}; {\bf (2)} altering the metric as inner product; and {\bf (3)} integrating adaptive early termination (Algorithm \ref{alg:search}).

\noindent {\bf Entry Points Selection.}
Unlike the random initialization used in \textsf{GNNS}, we generates navigation nodes from \textsf{SN}. For a query $q$, we first identify the closest cluster in the \textsf{SN}-\textsf{IVF} with cosine similarity, then randomly fills the initial pool from the corresponding inverted list. This is efficient because of the limited number of clusters, ensuring a swift and precise start to the search.

\noindent {\bf Altering the Metric.} Leveraging Theorem \ref{theorem:global} and \ref{theorem:local}, IP metric is enabled for \textsf{GNNS} on the \textsf{PSP}. Meanwhile, the max-heap and min-heap are interchanged adaptively due to the pursuing of answers with larger IP distances instead of smaller $L_2$ distances.  

\setlength{\textfloatsep}{0pt}
\begin{algorithm}[t]

  \caption{\textsc{Search On PSP}}
  \label{alg:search}
  \LinesNumbered
  \KwIn{Dataset $\mathcal{D}$, \psp $G$, query $q$, adaptive early termination function $ET(\cdot)$, candidate set size $l_s$, result set size $k$, query scale factor $\mu$.}
  \KwOut{Top $k$ result set $R$.}
  $R \gets \emptyset$; $Q \gets \emptyset$; $C \gets$ closest navigation cluster look-up;\\
  $P \gets$ random sample $l_s$ nodes from $C$; \\
  \For{each node $p$ in $P$}{
    $Q$.add($p$,$\langle p, q\rangle$);
  }
  $Q$.make\_max\_heap(); $R$.init\_min\_heap() \Comment{compare IP value}\\ 
  \While{$Q$.size()}{
    $p\gets Q.pop()[0]$; $R.insert((p, \langle p, q\rangle))$\\
    $f_p \gets$ get\_features($p$); \Comment{calculate features for $ET(\cdot)$} \\
    \If{$ET(f_p)$}{break; \Comment{$ET(\cdot)$ is true for stop}}
    \If{visited($p$)}{continue;}
    $N_p \gets$ neighbors of $p$ in $G$; $Q\gets$ batch\_insert($Q,N_p$);\\
    $Q.resize(l_s)$;$R.resize(k)$; \Comment{delete small-IP value nodes}\\
  }

  \textbf{return} $R$
\end{algorithm}
\setlength{\textfloatsep}{12pt plus 2pt minus 2pt}

\begin{figure}[tb!]
\centering
\centerline{\includegraphics[width=0.9\linewidth]{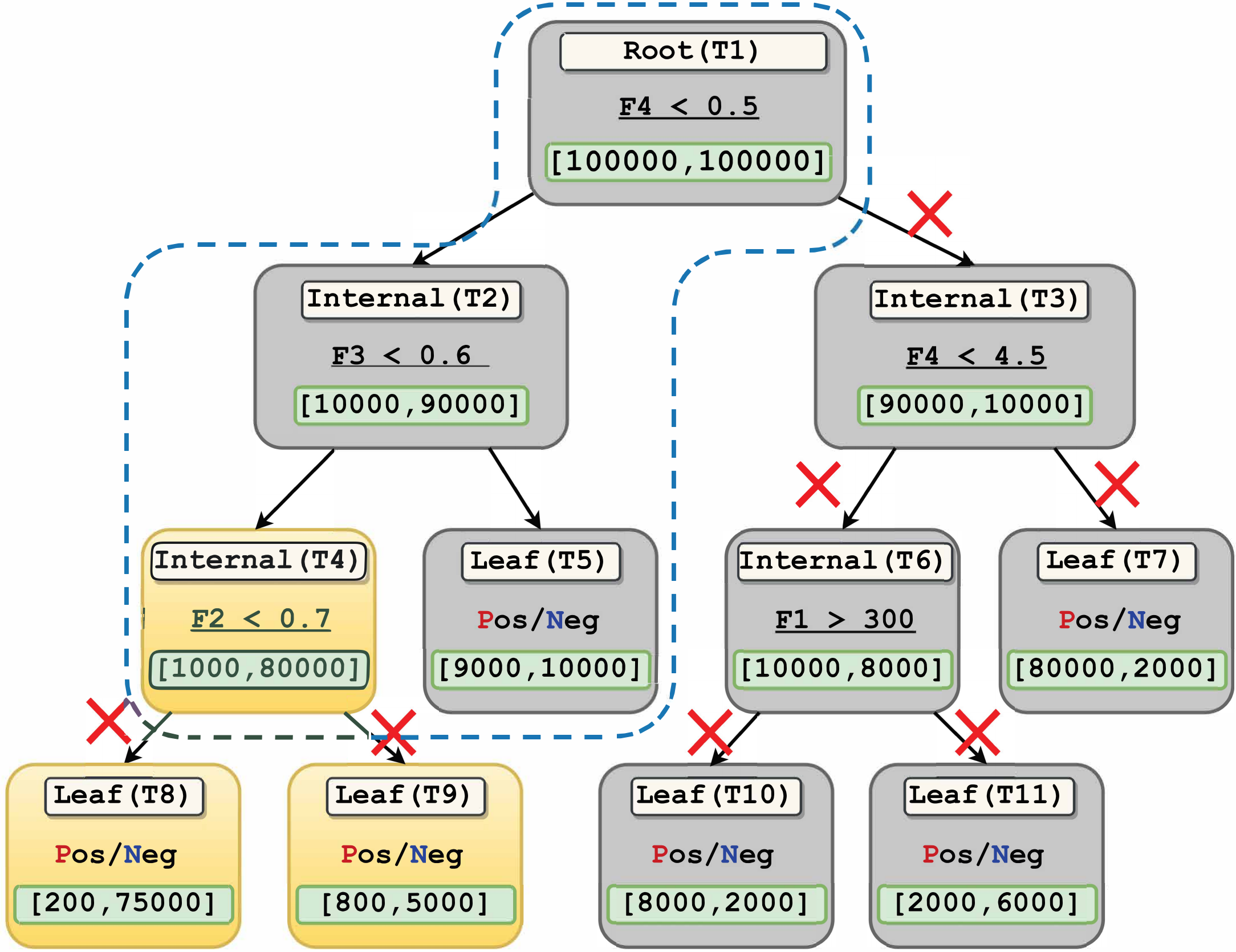}}
\caption{A showcase of a trained decision tree. The crosses denote pruned branches. The yellow nodes predict "stop". The conditional expressions are learnt decision logic based on selected features. "Pos/Neg" denotes the number of positive/negative samples fall in leaves, indicating stop tendency.}
\label{fig:decsion-tree}
\end{figure}

\noindent {\bf Adaptive Early Termination.} Early termination is crucial to search efficiency~\cite{liu2021ei,li2020improving,zhang2023vbase}. The exploration process is highly sensitive to the query distribution, yet standard \textsf{GNNS} is unaware of different query distributions and the accuracy of the result set. 
As shown in Figure \ref{fig:intuition-early-stop}(a), the distribution of the number of nodes visited exhibits a long tail, with most queries reaching 99\% recall with a small exploration depth. 
Instead of a uniform configuration for search parameters, adaptively stopping the search earlier for simpler queries can significantly enhance the amortized processing time and reduce excessive explorations. 
To equip \textsf{IP-GNNS} with "self-awareness", our method integrates a decision function to monitor key features indicative of the optimal stopping point, namely Adaptive Early Termination (\textsf{AET}), incorporates the following:

\eetitle{Feature Selection}. Features for the decision model are chosen based on the following principles. \textbf{(1) Relevance}: Features must be closely related to the \textsf{IP} metric, the query, and the tendency toward termination. 
\textbf{(2) Distinguishability}: Features must clearly distinguish between states before and after the termination point. 
\textbf{(3) Simplicity}: The computational cost of the features should be minimized to maintain efficiency. Theorem \ref{theorem:k-mip-neighbor} indicates the \textbf{vector norms}, the \textbf{IP values}, and \textbf{exploring neighborhood} of top-1 \mips solution are key aspects in feature engineering. Notably, the update frequency of the top-k candidate pool serves as a good indicator of the neighborhood exploration.
After extensive and rigorous experiments (Figure \ref{fig:intuition-early-stop}(b)), we finalized the feature set, detailed in Table \ref{tab:features}.
\begin{table}[tb!]
  \caption{Features used in adaptive early termination. EMA means exponential moving average.}
  \label{tab:features}
  \begin{tabular}{lc}
    \toprule
    Feature & Description \\
    \midrule
    F1: \textsf{IP} value  & EMA of (\textsf{IP} distance to $q$)\\
    F2: Norm/Min Norm & EMA of (norm / historical min norm) \\
    F3: \textsf{IP}/Max \textsf{IP}&  EMA of (ip / historical max ip) \\
    F4: Update Times & Top k update frequency (using EMA) \\
    \bottomrule
  \end{tabular}
\end{table}

\eetitle{Decision Model}. Utilizing the identified four highly discriminative features, a simple decision tree (\textsf{DT}) is employed, trained on labeled data points collected via searching on \psp. 
The data is generated through the following steps: {\bf (1)} Divide the base vectors into new base and new query sets; 
{\bf (2)} Perform searches on the new queries using the base indexed by \psp. For each query, record the node ID (boundary node) when the recall of the result set ceases to increase; 
{\bf (3)} Randomly sample nodes before the boundary node (labeled as 1) and after the boundary node (labeled as 0) to form the training data. 
{\bf (4)} Train the \textsf{DT} on the collected training data.

\eetitle{Function Generation}. To facilitate efficient implementation of the \textsf{AET} decision function, it is essential to convert the \textsf{DT} into C++ code. We propose a structured approach to simplify the \textsf{DT} while maintaining its effectiveness and efficiency: 
{\bf (1)} Limit the height of the \textsf{DT} to the number of features to prevent over-fitting. 
{\bf (2)} Adjust the prediction mechanism of the leaf nodes (Figure \ref{fig:decsion-tree}). A leaf node predicts "stop" only when the ratio $\frac{\# Negative\, Samples}{\# Positive\, Samples} > \theta$. This approach prioritizes higher confidence in stopping the search, reducing the risk of prematurely terminating. Users can adjust $\theta$ to control the aggressiveness of the \textsf{AET} function. 
{\bf (3)} Remove subtrees whose leaf nodes share the same predictions to simplify the tree. 
{\bf (4)} Translate the resulting \textsf{DT} into 'if-else' clauses in C++ from the tree structure. Figure \ref{fig:decsion-tree} is an example of this process. The final interpreted clause is "if (F4<0.5 \& F3 < 0.6) stop;else continue;"

\eetitle{Remarks.} Both candidate size $l_s$ and early-stop features affect the search recall by pruning the search space. Considering the monotonicity of these indicators w.r.t. search recall, one can tune the candidate size or the decision boundaries of the leaf nodes manually to get different recall, free of retraining the \textsf{DT} for each recall level.

\subsection{Complexity Analysis} 

\noindent \textbf{The Indexing} algorithm consists of two main stages: {\bf (1)} \textsf{NSSG} indexing and {\bf (2)} the \textsf{EF} and \textsf{SN}. 
Let $n$ denote the dataset size, $r$ the max degree of the graph, $d$ the dimension, $c$ the number of clusters in \textsf{SN}, and $m$ the number of navigation nodes in \textsf{SN}. 
The time complexity of \textsf{NSSG} indexing has an empirical complexity of $O(dn\log n)$~\cite{fu2021high}. 
For each node, the \textsf{EF} phase involves \textsf{IP} calculation ($O(ndr^2)$), sorting ($O(nr^2logr)$) and new neighbor selection ($O(nr^2)$), cumulatively yielding $O(nr^2(d+logr))$.
The \textsf{SN} phase, dominated by the $k$-means cluster assignment, scales as $O(ncd)$. Absorbing smaller constants into $c_4$, the overall indexing complexity of \psp is $O(c_4n\log n +n)$, dominated by $O(n\log n)$ term. 

\noindent \textbf{The Search} complexity for generally favored top-k \mips is approximated by $O(c_0\frac{\log n + c_1d}{\log R + c_2d} + c_3f(n)kRn^{1/d}\log n)$ in \S\ref{sec:analysis}. It is mainly dominated by the $O(kRn^{1/d}\log n)$ term. 
Empirical evaluations (detailed in \S\ref{sec:exp}) aligns with our estimation.

\noindent \textbf{Worst Cases.} The worst case will invalidate any acceleration techniques for approximate methods, e.g., points on a straight line. In such cases, the search complexity on \psp will downgrade to near $O(n)$. The indexing complexity of \psp will downgrade to $O(n^2)$.
\section{Experimental Evaluation}
\label{sec:exp}

\begin{table}[tb!]
  \caption{Dataset statistics. Dim indicates vector dimension.}
  \label{tab:prop}
  \resizebox{0.95\linewidth}{!}{
  \begin{tabular}{ccccc}
    \toprule
    Dataset & Base Size & Dim & Query Size& Modality\\
    \midrule
    \texttt{MNIST}~\cite{mnist} & 60,000 & 784 & 10,000 & Image\\
    \texttt{DBpedia100K}~\cite{oliva2001modeling} & 100,000 & 3072 & 1,000 & Text\\
    \texttt{DBpedia1M}~\cite{oliva2001modeling} & 1,000,000 & 1536 & 1,000 & Text\\
    \texttt{Music100}~\cite{morozov2018non,hu2008collaborative} & 1,000,000 & 100 & 10,000 & Audio\\
    \texttt{Text2Image1M}~\cite{text2image} & 1,000,000 & 200 & 100,000  & Multi\\
    \texttt{Text2Image10M}~\cite{text2image} & 10,000,000 & 200 & 100,000 & Multi\\
    \texttt{Laion10M}~\cite{laion} & 12,244,692 & 512 & 1,000 & Multi \\
    \texttt{Commerce100M} & 100,279,529 & 48 & 64,111 & E-commerce\\
    \bottomrule
  \end{tabular}}
\end{table}

We conduct extensive experiments on real-world datasets to validate the theoretical contributions of our proposed framework, as well as its efficiency, scalability, and practical applicability for the \mips task. Our analyses are guided by the following questions:

\begin{itemize}[left=0pt]
    \item \textbf{Q1}: How does the proposed theory align \mips with \nns ?
    \item \textbf{Q2}: How does \psp perform compared to established baselines across various modalities, dimensionalities, and cardinalities?
    \item \textbf{Q3}: Ablation study of the search efficiency of \psp.
    \item \textbf{Q4}: How is the viability of \psp in large-scale applications?
\end{itemize}

\subsection{Experimental Setup}
\label{sec:exp-setup}

\noindent {\bf Datasets.} All empirical evaluations were conducted on eight real-world datasets, covering diverse modalities, cardinalities and dimensionalities (Table \ref{tab:prop}): 
\textbf{\textsf{MNIST}}~\cite{mnist} and \textbf{\textsf{Music100}}~\cite{morozov2018non,hu2008collaborative} datasets are the established benchmarks for the \mips problem. \textbf{\textsf{DBpedia}}~\cite{oliva2001modeling} is extracted by \textsf{OpenAI} text-embedding-3-large model. \textbf{\textsf{Text2Image}}~\cite{text2image} is a cross-modality dataset where image embeddings (base) are extracted from Se-ResNext-101 and text embeddings (query) from a \textsf{DSSM} Model. \textbf{\textsf{Laion10M}}~\cite{laion} is a cross-modal retrieval dataset. The embeddings are generated by the CLIP-ViTB/32 model.
\textbf{\textsf{Commerce100M}}, generated from real traffic logs of a large-scale e-commerce platform. The embeddings are via an advanced method \cite{fu2024residual}. The dimension is set to 48 due to the application demands in industrial scenarios, while 16 to 64 are common choices in this field~\cite{ZhangWZTJXYY20,tian2102}, discussed in \cite{chen2025maximum}.

\noindent {\bf Competitors} were selected from advanced in-memory methods, encompassing \textbf{(1) \textsf{ip-NSW}}~\cite{morozov2018non}: A graph based method using inner-product navigable small world graph. \textbf{(2) \textsf{ip-NSW+}}~\cite{liu2020understanding}: An enhancement of \textsf{ip-NSW} that introduces an additional angular proximity graph. \textbf{(3) \textsf{M{\"o}bius-Graph}}~\cite{zhou2019mobius}: A graph based method that reduces the \mips problem to an \nns problem using M{\"o}bius transformation. \textbf{(4) \textsf{NAPG}} \cite{tan2021norm}: the state-of-the-art graph based method.  \textbf{(5) \textsf{Fargo}}~\cite{zhao2023fargo}: The latest state-of-the-art \textsf{LSH} based method with theoretical guarantees. \textbf{(6) \textsf{ScaNN}}~\cite{guo2020accelerating}: The latest state-of-the-art quantization method, an optimised version based on SOAR \cite{sun2024soar}. 

\noindent {\bf Implementation.} All baselines except for {\em ScaNN} are written in C++. The {\em ScaNN} library is called by Python bindings yet written in C++. The experiments are executed on a CentOS machine with 128G RAM on Intel(R) Xeon(R) CPU E5-2650 v4 @ 2.20GHz CPU. We use {\sf OpenMP} for parallel index construction, utilizing 48 threads for all methods. For query execution, we turn off additional optimizations.

\begin{table}[tb!]
  \caption{The average overlap ratio of \textsf{NNS}-\textsf{MIPS} search paths and recall@100 with varying $\mu$, on \textsf{MNIST}.}
  \label{tab:overlap}
  \resizebox{0.95\linewidth}{!}{
  \begin{tabular}{cccccc}
    \toprule
    Metric & $\mu=0.1$ & $\mu=1$ & $\mu=10$ & $\mu=100$ & $\mu=1200$ \\
    \midrule
    \texttt{Overlap ratio} & 7.4\% & 15.3\% & 72.4\% & 99.5\% & 100\% \\
    \texttt{Recall@100} & 0.0 & 0.08 & 0.83 & 1.0 & 1.0 \\
    \bottomrule
  \end{tabular}}
\end{table}

\noindent {\bf Parameter Settings.} \psp required tuning several hyper-parameters: number of neighbors in $k$\textsf{NN} graph ($K$), \textsf{NSSG} candidate size ($L$), maximum out-degree ($R$), minimal angle between edges ($\alpha$), and the number of nodes added in \textsf{EF} ($S$). By default, we set $K=400$, $L=800$, $R=40$, $\alpha=60^\circ$, and $S=5$ for all datasets. For a fair comparison, we also employ grid search to optimize the parameters of baseline methods over all datasets.

\noindent {\bf Evaluation Metrics.} For search performance comparison, we measure effectiveness and efficiency using two popular metrics: \textbf{(1) Recall vs. Queries Per Second (QPS)}, denoting the number of queries that an algorithm can process per second at each $recall@100$ level; and \textbf{(2) Recall vs. Computations}, indicating the number of distance computations performed by the search algorithm. Let $R_t'$ denote the set of $k$ vectors returned by the algorithm, and $R_t$ represent the ground truth, the recall@k is formally defined as 
$recall@k = \frac{|R_t \cap R_t'|}{|R_t|} = \frac{|R_t \cap R_t'|}{k}$
Additionally, we consider construction time and index size to evaluate the scalability in practice.

\noindent {\bf Experimental Results}
\label{sec:exp-result}
\noindent {\bf Exp-1: Theory Verification (Q1).}  To explore the \mips-\nns equivalence with respect to $\mu$, we conduct \textsf{GNNS} on an ideal \psp index, obtained by setting $L$ to the size of the base vectors and $R$ to infinity in Algorithm~\ref{alg:index}. This experiment is conducted on \textsf{MNIST}, and we also evaluate Theorem~\ref{theorem:k-mip-neighbor} on synthetic datasets sampled from a standard normal distribution, examining the relationship between top-k \mips solutions and the top-1 solution's Euclidean neighbors.

\eetitle{Duality of \textsf{NNS} and \textsf{MIPS} on Proximity Graph.} We assessed the average overlap ratio of search paths for \nns and \mips, and recall@100 for \mips, across all queries on \textsf{MNIST} under different $\mu$ (Table~\ref{tab:overlap}). The result indicates that a sufficiently large $\mu$ can effectively unify the \mips and \nns paths for all queries. Even sub-optimal $\mu$ yields reasonably good candidates due to highly overlapping paths. Case study in \cite{chen2025maximum} visualizes this. Additionally, we evaluate the overlap ratio between the top-k \mips solutions and the neighborhood of the top-1 \mips solution. The results in \cite{chen2025maximum} confirm that top-k \mips solutions are likely to be distributed in the Euclidean neighborhood of the top-1 \mips solution.

\begin{figure*}[tb!]
\centering
 \subfigure{\label{legend-qps-query}{
	\includegraphics[width=0.8\linewidth]{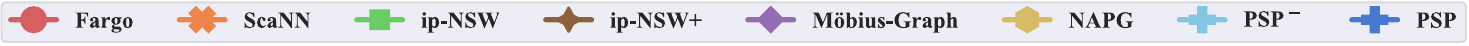}}}\\
\centerline{\includegraphics[width=0.94\linewidth]{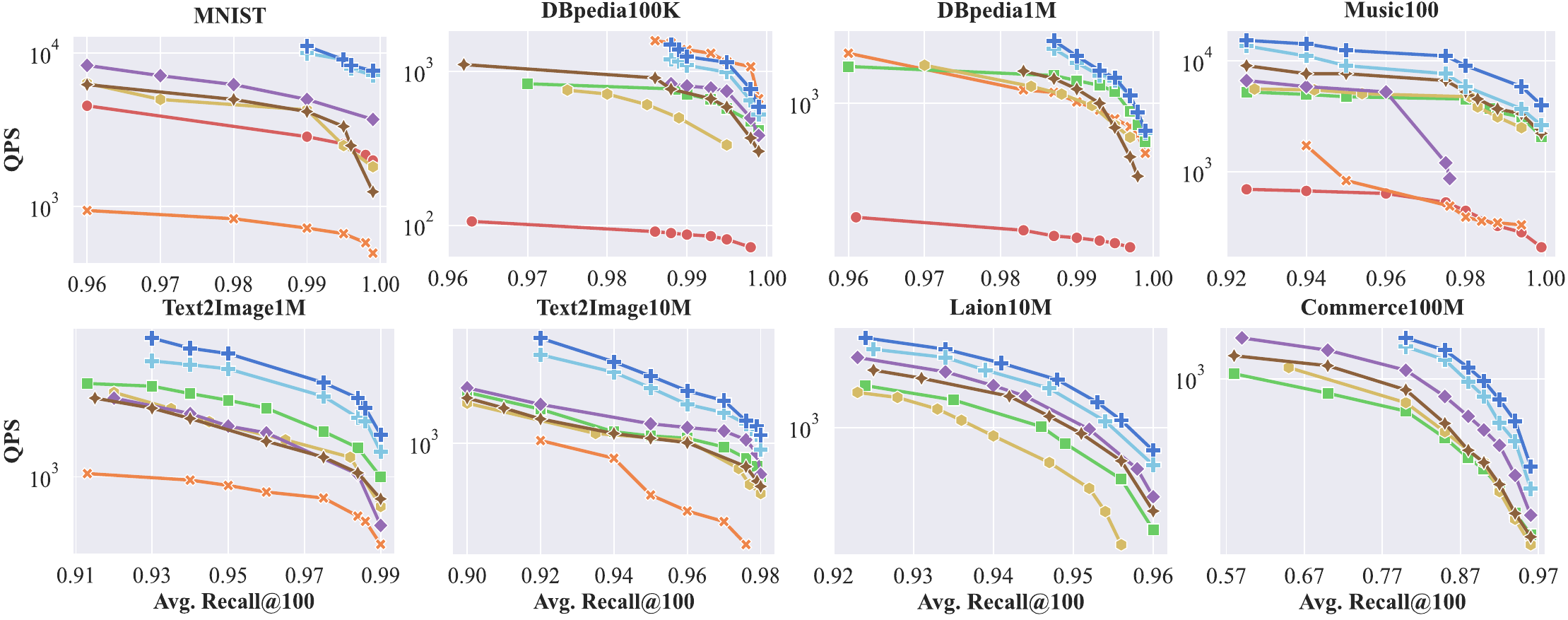}}
\centerline{\includegraphics[width=0.94\linewidth]{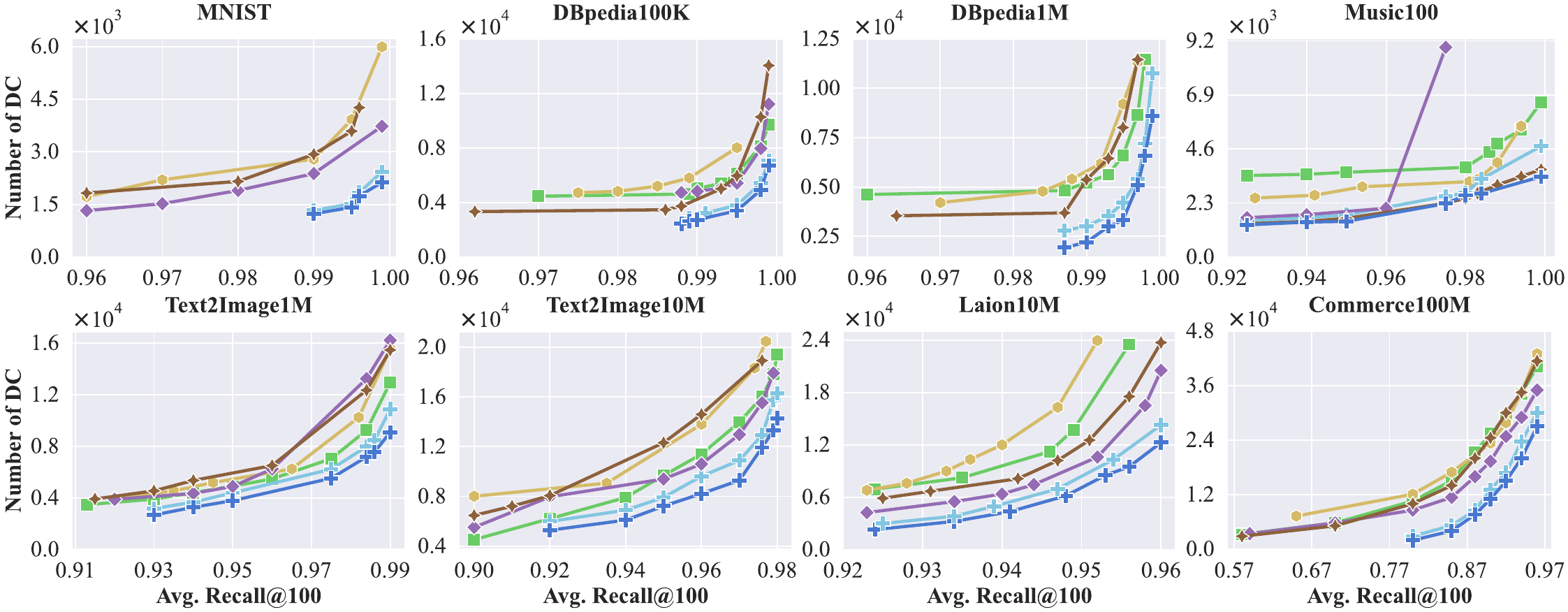}}
\caption{Experimental results of query process performance on eight datasets. DC denotes Distance Computation.}
\label{fig:exp-qps}
\end{figure*}

\begin{figure*}[tb!]
\centering
 \subfigure{\label{legend-qps-index}{
	\includegraphics[width=0.8\linewidth]{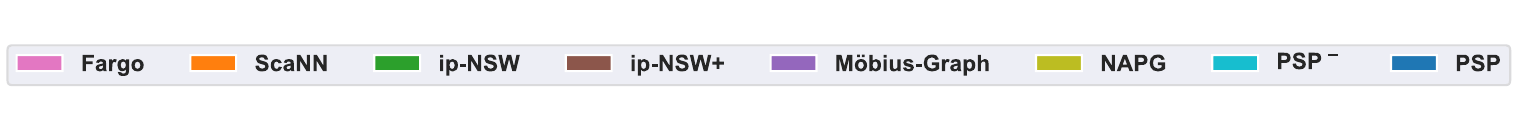}}}\\
\centerline{\includegraphics[scale=0.68]{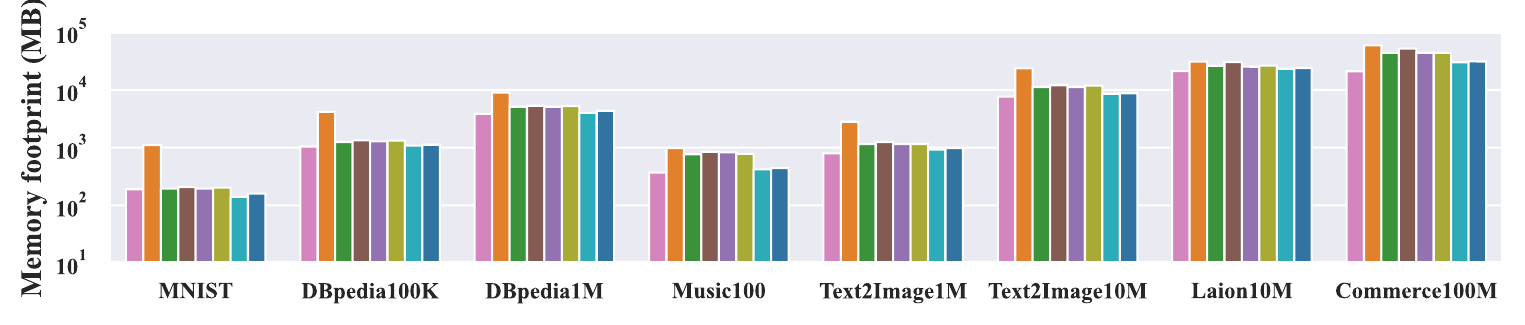}}
\centerline{\includegraphics[scale=0.68]{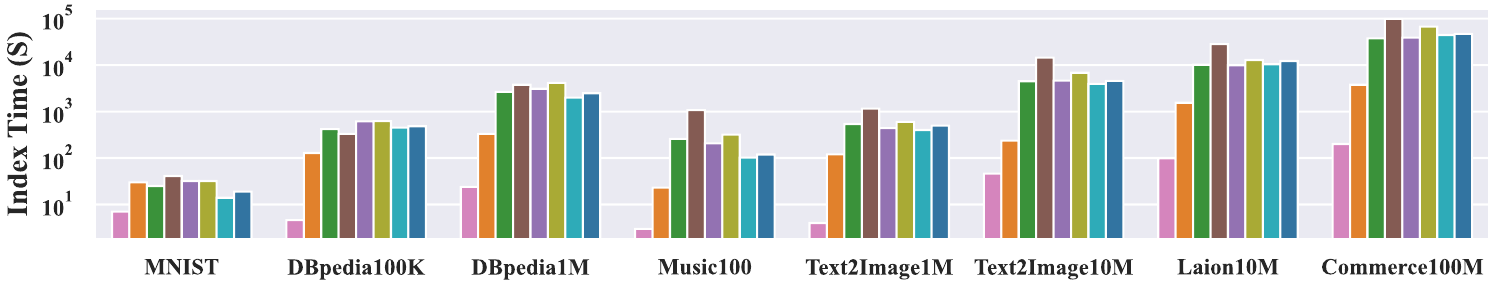}}
\caption{Experimental results on indexing time and memory footprint of different methods across eight datasets.}
\label{fig:memory}
\end{figure*}

\noindent {\bf Exp-2: Performance Assessments (Q2).} The results are presented in Figure \ref{fig:exp-qps} and \ref{fig:memory}. {\em The absence of methods in the figures} indicates either high processing time or low recall.

\eetitle{Query Processing Performance}. Figure~\ref{fig:exp-qps} presents the query processing performance for different methods. The top rows depict QPS at varying recall levels, where upper right is better. The bottom rows show the number of distance computations required, with lower right being better. Log scales are applied for wide value ranges.

\textbf{\textsf{PSP}} consistently outperforms all baselines across all datasets, achieving up to 37\% speedup over \textsf{M{\"o}bius-Graph} ($2^{nd}$ best) on \textsf{Laion10M}. This demonstrates \psp's superior efficiency with high-dimensional and large-scale data, leveraging its strong theoretical foundations and ensuring robust connectivity. In contrast, other graph-based methods suffer from connectivity issues, and non-graph methods' inefficiency mainly owes to inefficient indexing large-norm points.

\textbf{\textsf{PSP$^-$}}, representing an unoptimized \textsf{NSSG}, still surpasses other baselines across datasets by an average of 20\%-25\%, except on \textsf{Music100}. This aligns with our analysis that \textsf{MIPS} can be executed efficiently without transformation. Further optimizations, like \textsf{EF}, \textsf{SN}, and \textsf{AET}, improve performance by about 10\%-15\%.

\begin{figure}[t!]
	\centering
	\includegraphics[width=1\linewidth]{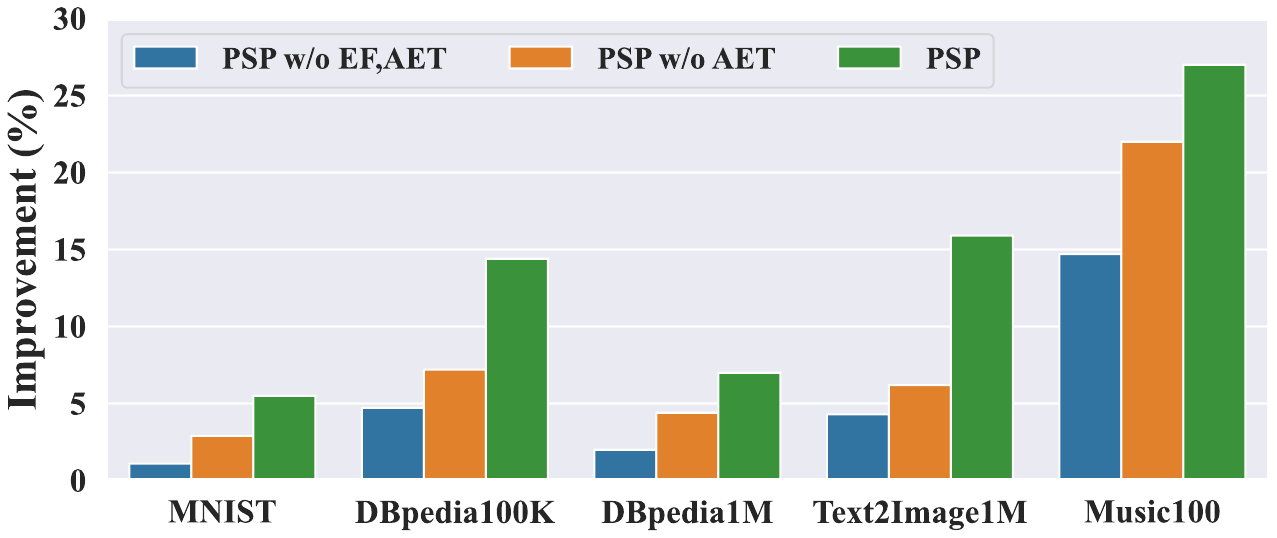}
	\caption{Ablation study results across five different scale datasets.  We measure the benefits of three optimization techniques on query execution.}
	\label{fig:exp-ablation}
\end{figure}

\textbf{Others:} \textbf{(1) \textsf{Fargo}} is inferior to those of graph-based algorithms, notably achieving only 90\% recall on \textsf{Commerce100M}. Its reliance on \textsf{LSH} and susceptibility to transformation-induced distortion limits its general performance. \textbf{(2) \textsf{ScaNN}} excels on \textsf{DBpedia100K} but struggles elsewhere, highlighting the sensitivity of quantization methods. For datasets not well-suited to quantization, such as large-scale or \textsf{IID} distributions, performance declines~\cite{li2019approximate}, as seen that its recall is limited on \textsf{Commerce100M}. \textbf{ (3) \textsf{ip-NSW} and \textsf{ip-NSW+}} lag behind \psp due to high graph density and precision limitations. Although \textsf{ip-NSW+} partially addresses precision issues, its performance can be inconsistent, especially on datasets like \textsf{Text2Image1M}, where the additional angular-graphs hurt performance. \textbf{ (4) M{\"o}bius-Graph} shows strong performance on \textsf{Commerce100M} but encounters precision bottlenecks on \textsf{Music100} and \textsf{DBpedia1M} due to violations of transformation-induced assumptions (\S\ref{sec:preliminaries}). \textbf{(5) \textsf{NAPG}} performs similarly to or worse than \textsf{ip-NSW} due to inefficient graph sparsification techniques and lack of theoretical support in their heuristics.

\eetitle{Indexing Performance.} Figure ~\ref{fig:memory} shows the index memory footprint and indexing time for various methods. \textsf{PSP} achieves a balanced memory footprint and indexing time across all datasets. On \textsf{Commerce100M}, \textsf{PSP} requires only 12 GB and 13 hours for indexing, making it feasible for daily updates—essential for adapting to evolving business metrics. In contrast, \textsf{ip-NSW+} takes over a day for indexing. While \textsf{Fargo} is the fastest and least memory-intensive, it suffers from poor query performance. Indices of \textsf{ScaNN, ip-NSW, ip-NSW+}, \textsf{Möbius-Graph}, and \textsf{NAPG} consume 1.5 to 2$\times$ the memory of \textsf{PSP}, reducing their practicality. Graph-based methods have an index size about 3$\times$ larger than \textsf{PSP}, excluding data, due to high redundancy in their edges. \textsf{ScaNN} also requires extensive memory for large tree structures to improve search precision. Table~\ref{tab:index_graph_feature} illustrates the graph features of \textsf{PSP} across different datasets. Utilizing its theoretical foundations, \textsf{PSP} exhibits a low average out-degree, a moderate clustering coefficient, and strong connectivity.

\begin{table}[t]
  \caption{Graph feature statistics of \textsf{PSP} on different datasets.}
  \label{tab:index_graph_feature}
  \resizebox{1\linewidth}{!}{
  \begin{tabular}{c c c c c c c}
    \toprule
    \multirow{2}*{Datasets} & \multicolumn{2}{c|}{\textsf{Index degree}} & \multicolumn{2}{c|}{\textsf{Cluster Coeff}} & \multicolumn{2}{c}{\textsf{Shortest path}} \\
    \cline{2-7}
    
    ~ & Avg. & Std. & Avg. & Std. & Avg. & Std.\\
    \midrule
    \textsf{MNIST} & 29 & 1.7 & 0.09 & 0.002 & 4.4 & 0.72 \\
  
    \textsf{DBpedia100K} & 35 & 0.02 & 0.03 & 0.004 & 3.7 & 0.51  \\
 
    \textsf{DBpedia1M} & 35 & 0.26 & 0.03 & 0.006 & 4.8 & 0.53  \\
    
    \textsf{Music100} & 21 & 3.7 & 0.10 &  0.008 & 5.3  & 0.78  \\
  
    \textsf{Text2image1M} & 39 & 1.5 & 0.05 & 0.001  & 5.9 & 0.9  \\
    
    \textsf{Text2image10M} & 39 & 1.6 & 0.04 & 0.009  & 7.6 & 1.3  \\
    \textsf{Laion10M} & 20 & 6.32 & 0.11 & 0.03  & 7.1 & 1.6  \\
    \textsf{Commerce100M} & 30 & 5.2 & 0.04 & 0.009 & 8.9 & 1.49 \\
    \bottomrule
  \end{tabular}
  }
\end{table}

\begin{table}[tb!]
  \caption{The querying and indexing performance over various data scales on \textsf{Text2image10M} at 99\% recall.}
  \label{tab:scale-commerce}
  \begin{tabular}{cccc}
    \toprule
    Dataset Scale & Query Time & Indexing Time & Index Size \\
    \midrule
    \texttt{Text2image100K} & 0.13 (ms) & 62 (s) & 15 (MB) \\
    \texttt{Text2image1M} & 0.46 (ms) & 498 (s)  & 148 (MB) \\
    \texttt{Text2image10M} & 0.95 (ms) & 5672 (s) & 1520 (MB) \\
    \bottomrule
  \end{tabular}
\end{table}

\noindent\textbf{Exp-2 Summary.} \textsf{PSP}'s superiority is attributed to : strong graph connectivity, no transform in space, efficient edge pruning, and tailored adaptations for \textsf{MIPS}. Superior search performance and efficient indexing makes \textsf{PSP} well-suited for large-scale applications.

\noindent {\bf Exp-3: Ablation Study (Q3).} We validates the impact of key components of \psp across five datasets. Results are recorded at recall@100 level of 99\% (Figure~\ref{fig:exp-ablation}). Key findings include:

\eetitle{Effect of Spherical Navigation.} \textsf{SN} improves average performance by 6.2\% across five datasets by starting searches in high-norm areas, reducing redundant computations. On \textsf{Music100}, it provides a 15\% speedup, suggesting a highly biased distribution of \mips solutions.

\eetitle{Effect of Edge Refinement.} \textsf{EF} adds a 3.7\% average improvement, and 8\% on \textsf{Music100}, by enhancing connectivity and increasing graph degree slightly, especially towards high-norm regions. The "Spherical Highways" introduced by \textsf{EF} serve as shortcuts, improving navigation efficiency. The effects of \textsf{EF} and \textsf{SN} may overlap, reducing additional benefits when combined.

\eetitle{Effect of Adaptive Early Termination.} \textsf{AET} yields an average improvement of 5.1\%, mitigating redundant paths once the optimal set is found. It is particularly effective for queries following long-tailed distributions, with improvements of 10.6\% on \textsf{Text2Image1M} and 8.3\% on \textsf{DBpedia100K}. Further speedups are achievable via smaller $\theta$ setting (\S\ref{sec:search}) with minor acceptable recall loss.

\begin{figure}[t!]
	\centering
	\includegraphics[width=1\linewidth]{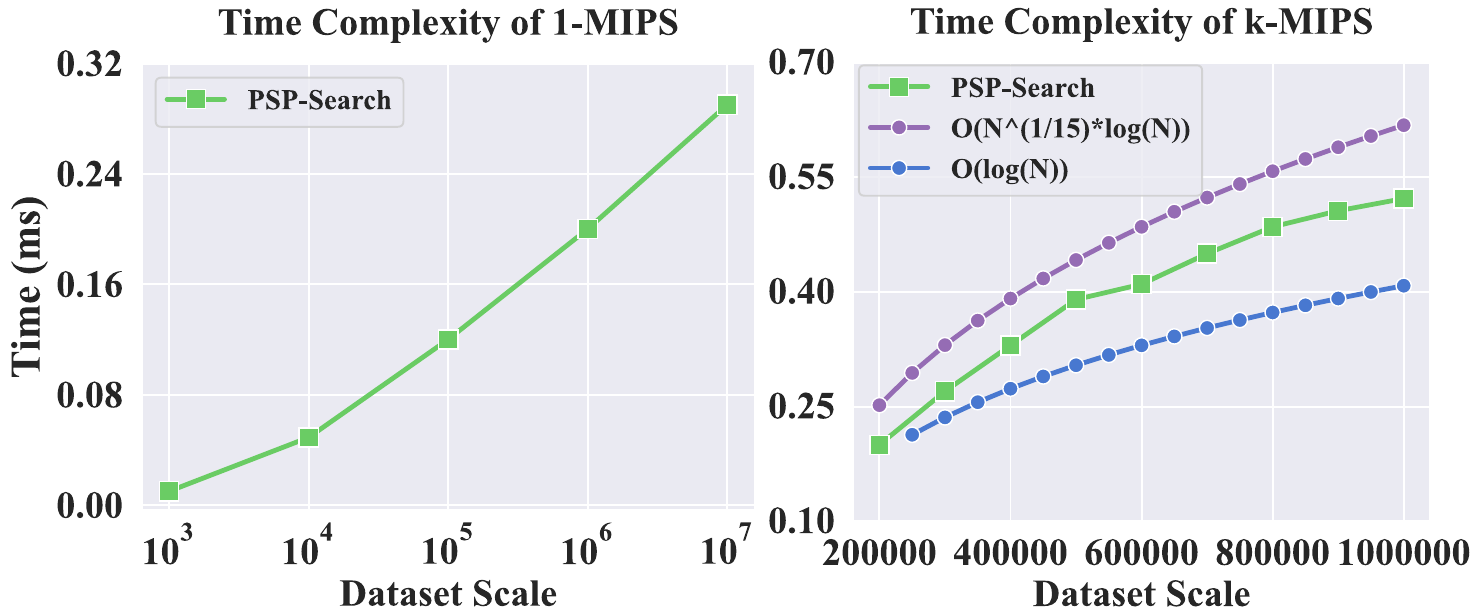}
	\caption{Query time versus data scale on \textsf{Text2image} at 99\% recall for top-1 and top-\textit{k} \mips. For top-1 queries, both the $x$ and $y$ axes use a logarithmic scale, resulting in an almost straight line, indicating a growth rate close to $O(\log n)$.}
	\label{fig:exp-complexity}
\end{figure}

\noindent {\bf Exp-4: Application Viability (Q4)} is determined by several aspects: 
the ease of hyper-parameter tuning, the scalability with dataset cardinality, the scalability with demanded answer number (k), the ease of tuning recall@k level, the consistency of query variance, and the flexibility to support diverse search scenarios. We find: 
\textbf{(1)} Extensive tests on \textsf{Text2Image} confirm the theoretical complexity estimates from \S\ref{sec:analysis}. Results (Figure~\ref{fig:exp-complexity}) show $O(\log n)$ scalability for top-1 \mips and near $O(\log n)$ scalability for top-k \mips, with manageable indexing times and approximately linear index size growth (Table \ref{tab:scale-commerce}). 
\textsf{PSP} achieves 99\% recall with a query time of just 0.9 ms on a 10 million scale dataset, demonstrating strong scalability with a small constant factor. 
\textbf{(2)} The performance of \psp is mainly determined by $R$ and quality of the base \textsf{NSSG}. Generally, higher quality of \textsf{NSSG} leads to better performance. Optimal $R$ exists yet remains consistent across all scales, allowing for efficient tuning on smaller subsets. \textbf{(3)} The query process time grows near sub-linearly with $k$, allowing for efficient search with large return quantity. \textbf{(4)} The query performance demonstrates strong robustness across different datasets (Table \ref{tab:varience}). The variance is small over random runs. \textbf{(5)} \textsf{PSP} can seamlessly support cosine similarity search (a special case of \textsf{MIPS}), allowing it to be widely adopted by various search vendors.

\begin{table}[tb!]
  \caption{Mean and standard deviation of the query execution time (ms) over various data scale at 95\% recall.}
  \label{tab:varience}
  \resizebox{1\linewidth}{!}{
  \begin{tabular}{cccc}
    \toprule
    Metric & Music100 & Text2image10M & Commerce100M \\
    \midrule
    time &  $0.095\pm 0.02$ & $0.543\pm 0.09$ & $2.631\pm 0.35$ \\
    \bottomrule
  \end{tabular}}
\end{table}


\section{Discussion}
\label{sec:discuss-limit}

\noindent \textbf{Strengths of New Dataset.} The new \textsf{Commerce100M} dataset is derived from large-scale real online traffic data collected from the Shopee e-commercial search. The vectors are generated with Resflow~\cite{fu2024residual}, advanced deep learning method in recommender and search. This dataset is unique for several reasons: \textbf{(1)} \textsf{Commerce100M}'s intrinsic dimension closely matches its actual dimension, leading to a distinct spatial topology compared to other "high-dimensional" datasets. \textbf{(2)} The queries represent users while the base vector represents groceries, a special type of cross-domain. \textbf{(3)} The performance on \textsf{Commerce100M} aligns perfectly with that on Shopee APP, rendering a real-world arena for this literature.

\noindent\textbf{Limitations.}
{\bf (1)} This study may not fully capture variability across extremely skewed norm distributions.
{\bf (2)} Our theory does not specifically prune redundant edges under \textsf{MIPS} setting. {\bf (3)} This work has not yet explored incremental indexing strategies.

\noindent\textbf{Future Works.} Building on the identified limitations, future research should focus on three key aspects: 
{\bf (1)} Ascertain the adaptability and robustness of our methods across a broader array of heterogeneous datasets. 
{\bf (2)} Consider developing acceleration techniques and theory specifically tailored to the \textsf{MIPS} problem.
{\bf (3)} Explore incremental indexing that can adapt real-time updates.

\section{Other Related Works}
\label{sec:related}

Inner Product (\textsf{IP}) is ubiquitous in metric learning, classification, clustering, knowledge graphs, recommender system, large language model, and information retrieval~\cite{yu2014large,fujiwara2023efficient,xu2020product,huang2020embedding,grbovic2018real,pan2023survey,zhao2023survey,wang2021milvus,guo2022manu, wang2024must}
\textsf{MIPS} methods can be categorized by their indexing strategies into \textsf{LSH}, tree, quantization, and graph based approaches:

\noindent {\bf LSH-based methods.} Adapting \textsf{LSH} to the \textsf{IP} metric includes L2~\cite{shrivastava2014asymmetric}, Correlation~\cite{shrivastava2015improved}, and the popular XBOX~\cite{bachrach2014speeding}, which introduces certain data distortion. Among the works based on XBOX~\cite{neyshabur2015symmetric,keivani2018improved,pham2021simple,yan2018norm,huang2018accurate,zhao2023fargo}, Fargo~\cite{zhao2023fargo} achieves state-of-the-art. 

\noindent {\bf Tree-based methods.} Early tree-based \textsf{MIPS} research~\cite{ram2012maximum,koenigstein2012efficient} struggled with high dimensionality. ProMIPS~\cite{song2021promips} tackles this issue by projecting the data into lower-dimensional spaces but it suffers from severe information loss. 
LRUS-CoverTree~\cite{ma2024reconsider} was designed to mitigate this, while it struggles with solving negative \textsf{IP} values. 

\noindent {\bf Quantization-based methods} target to accelerate distance computations with certain approximations. \textsf{ScanNN}~\cite{guo2020accelerating} combine the "VQ-PQ" framework with an anisotropic quantization loss, presenting a cutting-edge quantization-based solution. SOAR \cite{sun2024soar} uses
an orthogonality-amplified residual loss to enhance representations and reaches state-of-the-art and is part of \textsf{ScaNN} library.

\noindent {\bf Graph-based methods.} Proven effective for \textsf{NNS}~\cite{malkov2018efficient,fu2019fast,fu2021high,lu2021hvs,feng2023reinforcement,azizi2023elpis}, graph-based techniques have been adapted for \textsf{MIPS}. \textsf{ip-NSW}~\cite{morozov2018non} replaces Euclidean metric in \textsf{NSW}~\cite{malkov2014approximate} by \textsf{IP}, and \textsf{ip-NSW+}~\cite{liu2020understanding} further improves it by adding an angular proximity graph. \textsf{M{\"o}bius-Graph}~\cite{zhou2019mobius} introduces M{\"o}bius transformation, albeit with assumptions. 
Other algorithms ~\cite{tan2019efficient,tan2021norm} introduce heuristic edge selection strategies to accelerate the search.

\stitle{Other Acceleration Techniques.} ~\cite{li2018general} utilizes machine learning to optimize the \textsf{LSH} indices. EI-\textsf{LSH}~\cite{liu2021ei} enhances \textsf{LSH}-based methods with early termination techniques. Similar methodologies have also been explored in graph-based methods~\cite{li2020improving} for \textsf{NNS}.

\section{Conclusion}
\label{sec:concl}
In this paper, we align Maximum Inner Product Search (\textsf{MIPS}) with Nearest Neighbor Search (\textsf{NNS}) through robust theoretical foundations, presenting the first analytical framework for \mips efficiency on graph indices. 
To address the inherent challenges of biased solutions and excessive explorations in graph-based \mips, we propose a practical framework, featuring a novel graph index, \psp, and a novel lightweight adaptive early termination mechanism, \textsf{AET}. 
Extensive experiments validate the theoretical soundness, efficiency, scalability, and practical applicability of our approaches. 
Notably, our method has been deployed in the Shopee search engine, demonstrating superior performance. 
Additionally, we contribute the \textsf{Commerce100M} dataset to the public community, addressing the lack of e-commerce data in this domain.

\begin{acks}
This work was supported in part by the NSFC under Grants No. (62025206 and U23A20296), Zhejiang Province’s “Lingyan” R\&D Project under Grant No. 2024C01259, CCF-Aliyun2024004, Ningbo Yongjiang Talent Introduction Programme (2022A-237-G). Cong Fu is the main project advisor. Xiangyu Ke is the corresponding author of this work.
\end{acks}

\balance
\bibliographystyle{ACM-Reference-Format}
\bibliography{ref}

\newpage
\appendix
\section*{Appendix}
\label{sec-appendix}

\section{Complete Proof of THEOREM 2}

{\em Given a proximity graph $G=(V,E)$ and a query $q \in \mathbb{R}^d$, when using the standard Graph Nearest Neighbor Search (\gnns)~\cite{prokhorenkova2020graph} algorithm to decide the \mips solution for $q$, there exists a scalar $\bar{\mu}$ such that for $\forall \mu > \max(\bar{\mu},0)$, we can identify a node $p^* \in \mathcal{N}_o$ that satisfies: $p^* \in \left\{\argmax_{p \in \mathcal{N}_o}  \langle p, q \rangle\right\} \cap \left\{\argmin_{p \in \mathcal{N}_o} \delta(p, q')\right\}$. Here, $o$ can be any node on the search path of \gnns regarding the query $q'=\mu q$, and $\mathcal{N}_o=\left\{p|(p,o) \in E \right\}$ denotes $o$'s neighbor nodes.}

\begin{proof}
The search path of the Algorithm \ref{alg:gnns} on a proximity graph typically expands by selecting a node from the neighbors of the current node that minimizes the Euclidean distance to $q$ for \nns, yet maximizes the inner product for \mips (more details can be found in ~\cite{wang2021comprehensive}). To align the search paths under both metrics, we must unify the node-selection behavior along the path.

This challenge is addressed by localizing the problem: solving for the optimal $p^*$ among the neighbors at each greedy step, akin to the conditions established in Theorem \ref{theorem:global}. Let $l$ be the number of steps in the path. Define $U=\left\{\bar{\mu}_i| 1 \le i \le l\right\}$, the set of $\bar{\mu}$ solving all sub-problems. We have $\forall \mu > \sup \left(U\cup\{0\}\right)$, Algorithm \ref{alg:gnns} selects the same node under both \textsf{IP} and Euclidean metrics at each step.

In the special case where multiple neighbors qualify as local \mips solutions for $\mu q$, this ambiguity can be resolved by a simple adaptation to the standard Algorithm \ref{alg:gnns} protocol: always select the nearest neighbor of $\mu q$ with smallest \textsf{ID} among the qualified nodes.

In summary, there exists a scalar boundary $\sup \left(U\cup\{0\} \right)$ which forces the Algorithm \ref{alg:gnns} under both \textsf{IP} and Euclidean metrics to generate the same search path targeting the scaled query $\mu q$.
\end{proof}

\section{Complete Proof of THEOREM 3}

{\em  Consider a vector database $\mathcal{D} \subset \mathbb{R}^d$ containing $n$ points, where $n$ is sufficiently large to ensure robust statistical properties. 
    Let $\mathcal{G}$ be a SSG \cite{fu2021high} defined on $\mathcal{D}$. 
    Assume that the base and query vectors are independently and identically distributed (i.i.d.) and are drawn from the same Gaussian distribution, with each component having zero mean and a variance $\sigma^2$. 
    The expected length of the search path $L$ from any randomly selected start node $p$ to a query $q\in \mathbb{R}^d$ can be bounded by 
    $\mathbb{E}[L]<c_0\frac{\log n + c_1d}{\log R + c_2d}$,
    where $R$ is the max-degree of all possible $\mathcal{G}$, independent with $n$ \cite{fu2021high}, and $c_0,c_1,c_2$ are constants.}

\begin{proof}
Given Theorem \ref{theorem:global}, \ref{theorem:local}, along with the monotonic search property of \textsf{SSG} \cite{fu2021high}, Algorithm \ref{alg:gnns} identifies a greedy search path where each step minimizes the Euclidean distance to $\mu q$ while maximizes the {\sf IP} distance with respect to $q$, i.e. $\langle r_i, q\rangle > \langle r_{i-1}, q\rangle$. The expected length of the search path can be calculated as:
\begin{equation}
\nonumber
    \mathbb{E}[L] = \mathbb{E}_{p\in \mathcal{D}, q \in \mathbb{R}^d}\left[\frac{\langle r_{mip}, q\rangle - \langle p, q \rangle}{\mathbb{E}[\langle r_i, q \rangle - \langle r_{i-1}, q \rangle |r_{i-1}, q]}\right]
\end{equation}
where $r_{mip}$ is the \mips solution of $q$. 
The outer expectation cannot be simplified directly due to potential dependencies among $p$, $r_i$, and $q$. 
Given that the base and query vectors are finite and drawn from the same distribution, we can enclose them in a hypersphere of diameter $2H$~\cite{weisstein2002hypersphere}, which is independent of $p$, $r_i$, and $q$. 
Since $\langle p, q \rangle < H^2$ and $\langle r_{mip}, q \rangle < H^2$, we can derive:
\begin{equation}
\nonumber
    \mathbb{E}[L] < \frac{2H^2}{\mathbb{E}_{p\in \mathcal{D}, q \in \mathbb{R}^d}\left[\mathbb{E}\left[\langle r_i, q \rangle - \langle r_{i-1}, q \rangle |r_{i-1}, q\right]\right]}
\end{equation}

Although the exact solution to this inequality is intractable, a practical approach involves approximating both the numerator and the denominator. 
By applying Extreme Value Theory ({\sf EVT})~\cite{smith1990extreme}, we can derive a tight upper bound for this approximation, ensuring convergence to the bound as $n$ becomes sufficiently large.

As $H$ is the half-diameter, $H^2$ is the maximum squared norm of the dataset $\mathcal{D}$.
We define $M_0 = H^2 = \max_1^n\{X_1^2, ..., X_n^2\}$, where $X_i$ are element-wise i.i.d. samples from $\mathcal{N}(0,\sigma^2)$.
Consequently, $X^2$ follows a chi-square distribution with $d$ degrees of freedom. 
The Moment Generating Function ({\sf MGF})~\cite{curtiss1942note} of $M_0$ is given:
\begin{equation}
    \nonumber
    MGF(t)_{X^2} = (1-2\sigma^2t)^{-\frac{d}{2}}, 0< t < \frac{1}{2\sigma^2}
\end{equation}
From Jensen's Inequality~\cite{mcshane1937jensen} with $\phi(x)=e^{t_0x}, t_0>0$, we have:
\begin{equation}
    \nonumber
    \begin{aligned}
        e^{t_0\mathbb{E}[M_0]} &\le \mathbb{E}[e^{t_0M_0}] = \mathbb{E}\max_{i=1}^ne^{t_0X_i^2} \\
        &\le \sum_1^n\mathbb{E}\left[e^{t_0X_i^2}\right] = nMGF(t_0)_{X^2} 
    \end{aligned}
\end{equation}
\begin{equation}
    \nonumber
    \mathbb{E}[M_0] \le \frac{1}{t_0}\left(\log n + \frac{d}{2}\log (1-2\sigma^2t_0)^{-1}\right) 
\end{equation}
Here, $t_0$ is a hyper-parameter that can be optimized within the range $(0, 0.5)$ to enhance the tightness of this bound. When $n$ is sufficiently large, this upper bound serves as a good approximation of $\mathbb{E}[M_0]$.

A similar approach can be applied to the random variable $XY$ for approximating the denominator of Eq.~\ref{eq:ubl}, where $X$ and $Y$ element-wise i.i.d. sampled from $\mathcal{N}(0,\sigma^2)$. 
In this case, $XY$ follows a distribution characterized by a modified Bessel function of the second kind~\cite{bowman1958introduction}. The {\sf MGF} for $XY$ is given by:
\begin{equation}
    \nonumber
    MGF(t)_{XY} = \left(\sqrt{1-\sigma^2t_1^2}\right)^{-d}
\end{equation}

Define $M_1 = \max_0^R\{X_1,..,X_R\}$, where $R$ is the maximum degree of \textsf{SSG} $\mathcal{G}$. Applying Jensen's Inequality again, we obtain:
\begin{equation}
    \nonumber
    \begin{aligned}
        e^{t_1\mathbb{E}[M_1]} &\le E[e^{t_1M_1}] \\
        \mathbb{E}[M_1] &\le \frac{1}{t_1}\left(\log R + d \log (1-\sigma^2t_1^2)^{-1}\right)
    \end{aligned}
\end{equation}
where $0<\sigma^2t_1^2<1$. 
This derived upper bound assumes a distribution centered at the origin. 
By applying Re-scaling (shifting), we can derive the bound for the case of neighbors of $r_{i-1}$ (points centered at $r_{i-1}$ in a Euclidean proximity graph) as follows:
\begin{equation}
    \nonumber
\begin{aligned}
    \mathbb{E}[\langle r_i,q \rangle] &= \mathbb{E}\left[\max_{m=1}^{R}\{\langle r_m,q \rangle|r_m \in N(r_{i-1})\}\right] \\
    &\le \mathbb{E}[\langle r_{i-1}, q \rangle] + \frac{1}{t_1}\left(\log R + d \log (1-\sigma^2t_1^2)^{-1}\right)
\end{aligned}
\end{equation}

Substituting them into Eq.(\ref{eq:ubl}), we have:
\begin{equation}
    \nonumber
\begin{aligned}
    \mathbb{E}[L] &< \frac{2H^2}{\mathbb{E}_{p\in \mathcal{D}, q \in \mathbb{R}^d}\left[\mathbb{E}\left[\langle r_i, q \rangle - \langle r_{i-1}, q \rangle |r_{i-1}, q\right]\right]} \\
    &\approx \frac{\frac{2}{t_0}\left(\log n + \frac{d}{2}\log (1-2\sigma^2t_0)^{-1}\right)}{\frac{1}{t_1}\left(\log R + d\log (1-\sigma^2t_1^2)^{-1}\right)}
\end{aligned}  
\end{equation}
where $R$ is the maximum degree of $\mathcal{G}$, independent of $n$ and choices of $p$, $q$, and $r_i$\cite{fu2021high}.
Thus, the outer expectation can be eliminated. 
When $t_0$ and $t_1$ are optimized for the tightest approximation, introducing constants $c_0$, $c_1$ simplifies the formula to:
\begin{equation}
    \nonumber
\mathbb{E}[L] < c_0\frac{\log n + c_1d}{\log R + c_2d}
\end{equation}
\end{proof}

\section{Details of Commerce100M dataset.} \textsf{Commerce100M} is derived from real traffic logs of a large-scale e-commerce platform, with embeddings generated by an advanced, pre-trained personalized deep retrieval model. The embeddings have a dimensionality of 48, aligning with industrial norms and resource constraints, where vector dimensions typically range from 16 to 64 due to the vast number of candidates and memory limitations \cite{ZhangWZTJXYY20,tian2102}. In this setup, storing 100 million 48-dimensional vectors consumes about 20GB of memory, leaving 30GB for indices under a 50GB per-shard memory allocation for MIPS tasks. Additionally, these vectors are utilized for ranking items among retrieved candidates, with the ranking engine capping the dimension at 64 for efficient similarity ranking.

Query vectors represent joint user and search keyword profiles, while base vectors encompass over 100 million popular grocery items. The query vectors are categorized into three types. Each vector is composed of three segments, each containing 16 dimensions, used in different combinations for various similarity calculations reflecting user behaviors:

\begin{itemize}
    \item When using part1 (16 dimensions) of the query and part1 (16 dimensions) of the item for similarity calculation, it indicates the user's click tendency.
    \item When using the concatenation of part1 and part2 of the query with the concatenation of part1 and part2 of the item for similarity calculation (32 dimensional vector), it indicates the user's add-to-cart tendency.
    \item When using the concatenation of part1, part2, and part3 of the query with the concatenation of part1, part2, and part3 of the item for similarity calculation (48 dimensional vector), it indicates the user's purchase tendency.
\end{itemize}

Moreover, each query contains a sequence of items with which the user has actually interacted and their corresponding labels. A query may correspond to one or multiple interacted items. Different labels represent different levels of interaction:

\begin{itemize}
    \item label=3: The user only clicked.
    \item label=4: The user clicked and added to the cart.
    \item label=5: The user clicked, added to the cart, and made a purchase.
\end{itemize}

\stitle{Insights from deploying \textsf{PSP}:} Shopee's e-commerce retrieval application is driven by specific real-world business needs. \textbf{(i)} On the Shopee platform, a large number of products are batch-listed and delisted within a brief time frame. \textbf{(ii)} Considering the limitations of online service resources and the requirements for multi-replica disaster recovery and traffic distribution, we must constrain the index update time, the memory utilized for index updates, and the memory required for index serving. Many existing methods either fail to meet the retrieval efficiency requirement (e.g., \textsf{ScaNN}), or consume excessive index resources. In contrast, \textsf{PSP} perfectly matchs all our requirements.

\section{Additional results for Fargo and M{\"o}bius-Graph.} 
\begin{figure*}[htb!]
\centering
\centerline{\includegraphics[width=1\linewidth]{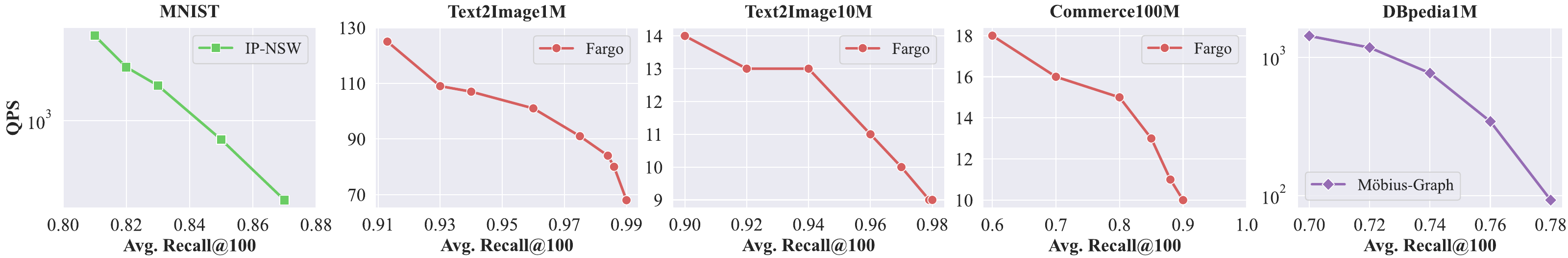}}
\vspace{-2mm}
\caption{Query performance of ip-NSW, Fargo and M{\"o}bius-Graph.}
\vspace{-1.2ex}
\label{fig:exp-fargo-mobius}
\end{figure*}

Figure~\ref{fig:exp-fargo-mobius} demonstrates the the query performance of \textsf{Fargo} on \textsf{Commerce100M} and \textsf{M{\"o}bius-Graph} on \textsf{DBpedia1M}, which is absent in Figure~\ref{fig:exp-qps} due to either excessively high processing time or low recall. Specifically, on the \textsf{Commerce100M} dataset, \textsf{Fargo} achieves only 90\% recall@100 and exhibits significantly slower performance. \textsf{M{\"o}bius-Graph}'s recall on \textsf{DBpedia1M} gets stuck at 78\% because its vector space transformation approach, which, as discussed in \S\ref{sec:preliminaries}, relies on assumptions that are frequently violated in real-world data, thereby impairing performance.

\section{Empirical Verification of Theoretical Results.}
\begin{figure*}[tb!]
\centering
\centerline{\includegraphics[width=1\linewidth]{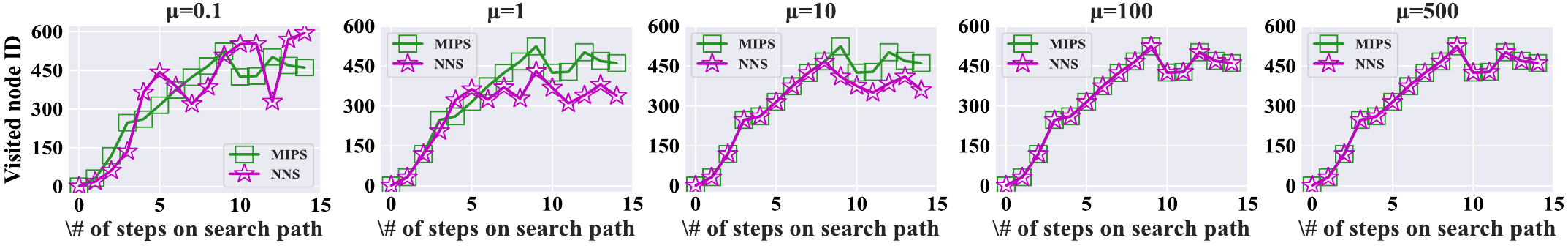}}
\vspace{-2mm}
\caption{A case study on the overlap of \textsf{MIPS}'s and \textsf{NNS}'s search paths with different configuration of $\mu$.}
\vspace{-1.2ex}
\label{fig:exp-duality}
\end{figure*}

\begin{figure*}[tb!]
\centering
\centerline{\includegraphics[width=1\linewidth]{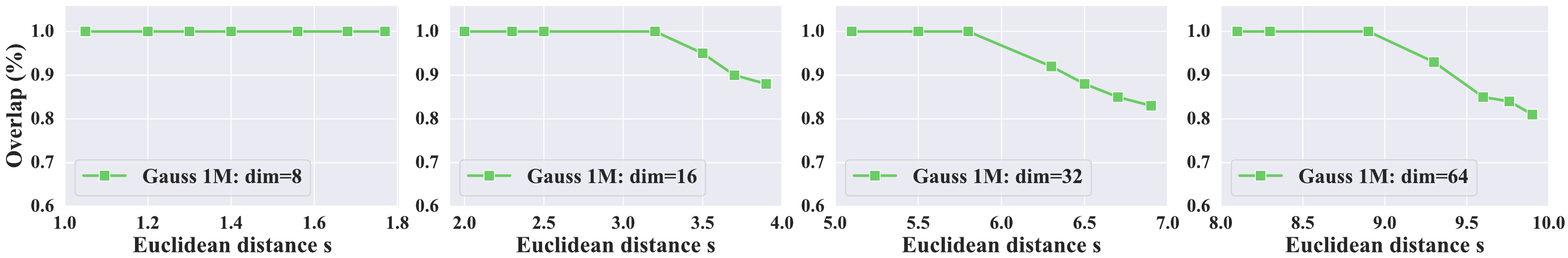}}
\vspace{-2mm}
\caption{A case study on the overlap of overlap between k-MIPS solutions and the neighborhood around 1-MIPS solution.}
\vspace{-1.2ex}
\label{fig:exp-gauss}
\end{figure*}
\stitle{Exp-1: Theory Verification (Q1).} As detailed in \S\ref{sec:analysis}, the key to equating \mips with \nns on a proximity graph is finding an appropriate query-scaling factor $\mu$. We explore this property by conducting \textsf{GNNS} on an ideal \psp index. An ideal \psp can be obtained by executing Algorithm~\ref{alg:index} with $L$ set to the cardinality of the base vectors and $R$ set to infinity. We perform this experiment on \textsf{MNIST}. Moreover, we evaluate Theorem~\ref{theorem:k-mip-neighbor} on synthetic datasets based on standard normal distribution, about the relationship between top-k \mips solutions and the Euclidean neighborhood around top-1 \mips solution.

\eetitle{Duality of \textsf{NNS} and \textsf{MIPS} on Proximity Graph.} The case study is conducted by: (1) Randomly select a query and run \textsf{GNNS} with varying $\mu$ values: $\mu$ in $\{0.1, 1, 10, 100, 500\}$; (2) Trace node IDs visited during the search up to a maximum of $15$ steps, as \textsf{MIPS} usually reaches its solution within this limit. The results, as shown in Figure \ref{fig:exp-duality}, demonstrate that the search paths of \mips and \nns diverge initially but increasingly overlap as $\mu$ increases. At $\mu=100$, the search paths coincide completely. Furthermore, we assessed the average overlap ratio of search paths for \textsf{NNS} and \textsf{MIPS}, and recall@100 for \textsf{NNS}, across all queries on \textsf{MNIST} under different $\mu$ (as shown in Table~\ref{tab:overlap}). The result indicates that while precisely defining the lower bound of $\mu$ for every possible query is impractical, a sufficiently large $\mu$ can effectively unify the \textsf{MIPS} and \textsf{NNS} paths for all queries. Sub-optimal $\mu$ does not lead to collapsed results, yet 
yields reasonably good candidates due to highly overlapped paths. 

\begin{figure}[t!]
	\centering
	\includegraphics[width=1\linewidth]{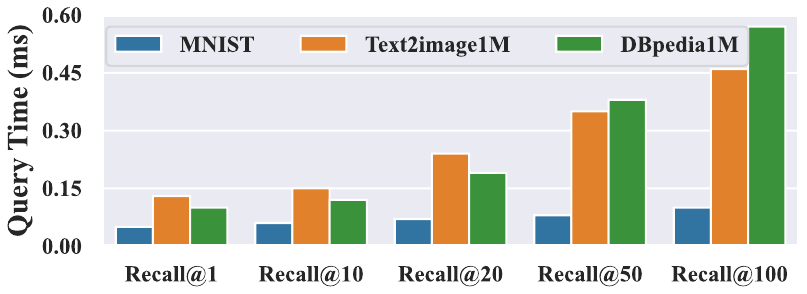}
		\label{fig:performance-with-vary-k}
	\vspace{-5mm}
	\caption{The query process performance of various k of recall on \textsf{MNIST}, \textsf{Text2image1M} and \textsf{DBpedia1M} at 99\% recall.}
	\label{fig:scalability-K}
\end{figure}

\eetitle{Overlap Between K-\mips and Nearest Neighbors.} Theorem \ref{theorem:k-mip-neighbor} reveals a high confidence of the overlap between top-k \mips solutions and the neighborhood around the top-1 \mips solution. We evaluate how the overlap ratio changes with $s$ on standard Gaussian distribution datasets ($\mathcal{N}(0,1)$). The results shown in Figure~\ref{fig:exp-gauss} verify our analysis that after the top-k \mips solutions are distributed around  the top-1 \mips solution's Euclidean neighborhood with a high confidence. With higher dimensions and larger Euclidean distance range, the overlap gradually goes down, which aligns with intuition and the curse of dimensionality. In higher dimensional spaces, the relationship between vectors are more complicated. The duality of Euclidean and IP metric will degrade. Luckily, most real-world datasets, though with high dimension, lie on a very low-dimension manifold. For example, the intrinsic dimension of 960-dimensional GIST1M dataset is about 20+, making the search on it as simple as on a Gaussian random dataset of 20 dimensions.

\begin{figure*}[t!]
        \vspace{1mm}
	\centering
	\includegraphics[width=\linewidth]{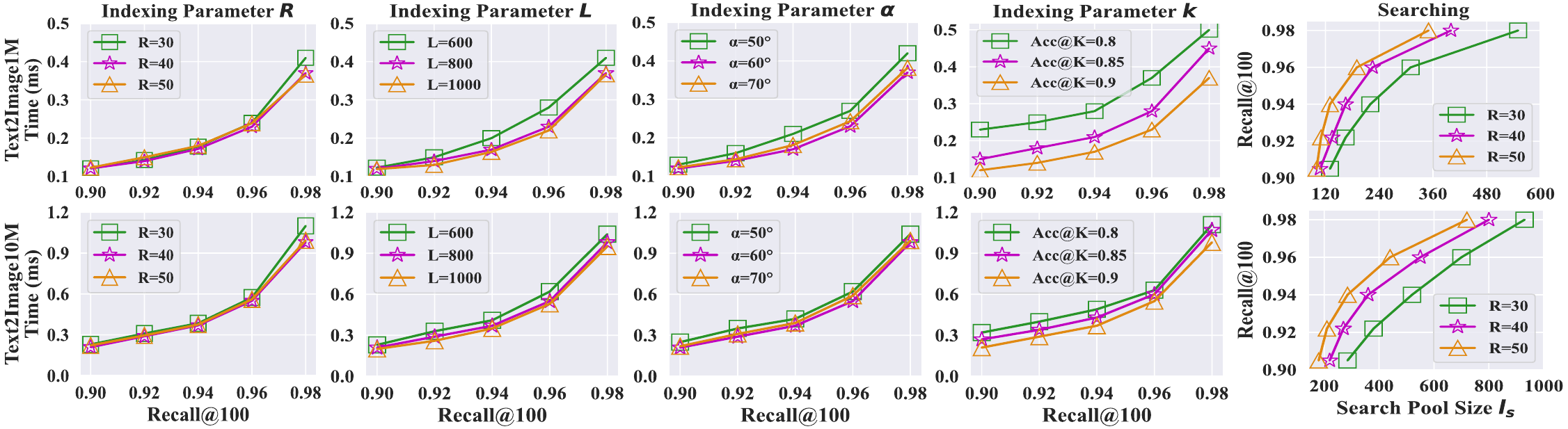}
	\vspace{-5mm}
	\caption{Parameter sensitivity experimental results for key parameters and factors on \textsf{Text2image1M} and \textsf{Text2image10M}}
	\label{fig:exp-parameters}
	\vspace{-3mm}
\end{figure*}

\section{Parameter sensitivity experiments.} 
\begin{figure}[tb!]
\centering
\centerline{\includegraphics[width=1\linewidth]{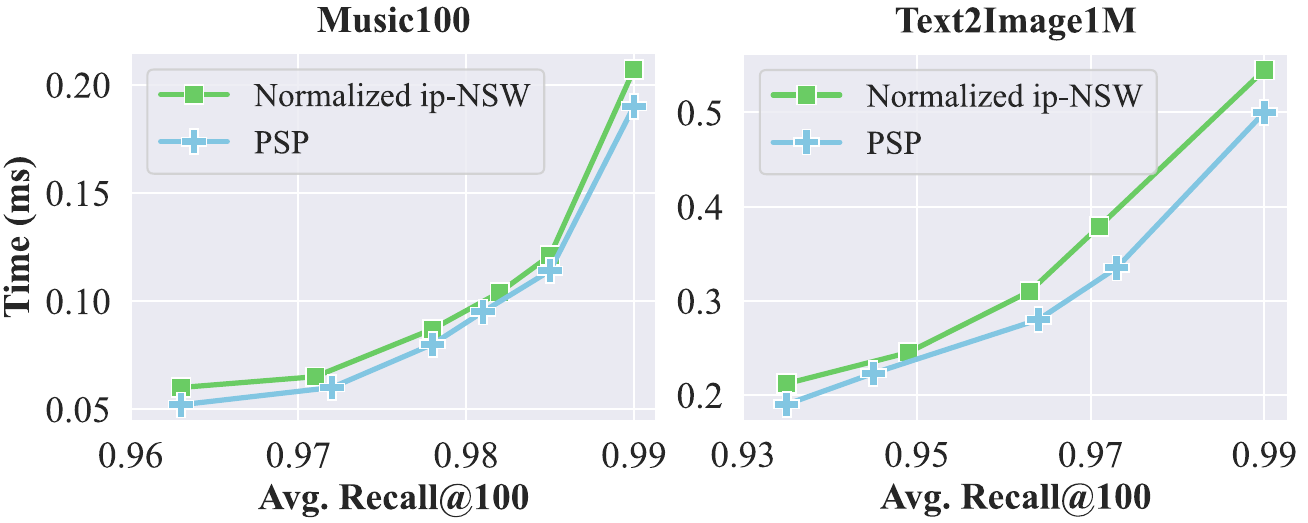}}
\vspace{-2ex}
\caption{Experimental results of query performance using cosine similarity on the Music100 and Text2image1M datasets.}
\vspace{-0.9ex}
\label{fig:exp-cos}
\end{figure}
\eetitle{Parameter Sensitivity.} 
Parameter tuning is often easier when patterns are predictable or when the parameter-performance relationship is smooth and convex. 
Our evaluations of \textsf{PSP}'s parameter sensitivity reveal several key {\em consistent patterns of optimal configurations}: 
{\bf (1)} Parameters like $R$ (maximum out-degree) and $S$ (number of nodes added in \textsf{EF}) remain consistent {\em across different dataset scales}, as depicted in Figure~\ref{fig:exp-parameters}. This consistency simplifies parameter tuning on a smaller subsets of data.
{\bf (2)} For the minimum angle between edges ($\alpha$), $60^\circ$ is optimal across all datasets, aligning with previous findings \cite{fu2021high}. 
{\bf (3)} For other parameters like $Acc@K$ (accuracy of the $k$-\textsf{NNG} to initialize \textsf{NSSG}) and $L$ (candidate size for \textsf{NSSG}), larger values generally enhance accuracy but potentially increase processing time. Users can adjust these parameters based on specific performance requirements and computational constraints.
\textbf{In summary}, parameter tuning for \psp indexing and searching is straightforward because of the predictable and consistent patterns observed across various datasets. 
The default configuration outlined in \S\ref{sec:exp-setup} ensures robust performance and is an alternative to parameter tuning, given \psp's demonstrated superiority across the evaluated datasets.

\section{Additional Scalability Experiments.}

\eetitle{Scalability with Recall Quantity (k).} To assess scalability relative to varying values of $k$, we adjusted $k$ from $\{1, 10, 20, 50, 100\}$ across the datasets \textsf{MNIST}, \textsf{Text2Image1M}, and \textsf{DBpedia1M}, each differing in cardinality and dimensionality. The results (Figure~\ref{fig:scalability-K}) suggest that both cardinality and dimensionality impact scalability in $k$. Higher cardinality and dimensionality both contribute to a faster increase in query processing time, aligning with expectations. Despite this, the scaling trend in query processing time remains practical due to the small basis of the time (lower than 1ms per query).

\eetitle{Scalability with Cosine Similarity Search.} We perform experiments on the \textsf{Music100} and \textsf{Text2image1M} datasets using cosine similarity search. We compared \textsf{PSP} with a state-of-the-art algorithm that first normalizes the dataset by scaling all data points to a norm of 1, constructs the graph using \textsf{ip-NSW}, and then applies cosine similarity search, referred to as \textbf{normalized ip-NSW}. In this case, cosine similarity, inner product, and nearest neighbor search are equivalent. As for \textsf{PSP}, We directly altered the metric to cosine similarity rather than normalizing the dataset. Figure~\ref{fig:exp-cos} demonstrates that \textsf{PSP} yields favorable results when applying cosine similarity to unnormalized datasets. We also directly tested \textsf{ip-NSW} for cosine similarity search but \textbf{encountered precision issues with \textsf{Music100} (70\%) and \textsf{Text2image1M} (89\%)}. This stems from its tendency to connect nodes with large norms, which are less useful for cosine similarity. This highlights the effectiveness of \textsf{PSP}.

\section{Implications and Open Questions}

\eetitle{Impact and Implication.} This study represents the first attempt to align \textsf{MIPS} and \textsf{NNS} without altering the origin vector space, the first theoretical framework on search efficiency for graph-based \mips methods, introducing a novel approach that could inspire a range of non-transformative methods. 
The theoretical insights provided have broad implications, extending beyond just graph-based vector retrieval to potentially influence other database and machine learning tasks.

\eetitle{Limitations.} 
{\bf (1)} The current study's experiments may not fully capture performance variability across extremely skewed data distributions, particularly with outliers or distant clusters of varying norms.
{\bf (2)} Our theory does not specifically address redundant computation pruning tailored for the \textsf{IP} metric. Developing solutions for this issue could further enhance the efficiency of our method. 
{\bf (3)} The research has not yet explored incremental indexing strategies.

\eetitle{Future Works.} Building on the current study's implications and its identified limitations, future research should focus on several key areas: 
{\bf (1)} Further investigations should ascertain the adaptability and robustness of our methods across a broader array of heterogeneous datasets. 
{\bf (2)} We consider developing acceleration techniques specifically tailored to the \textsf{MIPS} problem, backed by robust theoretical frameworks. 
{\bf (3)} There is a need to explore incremental indexing techniques that can accommodate real-time updates to data without requiring a complete rebuilding of the index.

\balance
\end{document}